\documentclass[10pt,twocolumn,twoside]{IEEEtran}
\def\ps@headings{%
\def\@oddhead{\mbox{}\scriptsize\rightmark \hfil \thepage}%
\def\@evenhead{\scriptsize\thepage \hfil \leftmark\mbox{}}%
\def\@oddfoot{}%
\def\@evenfoot{}}
\usepackage{epsfig, epsf, array, latexsym, graphics, multirow, color,enumitem}
\usepackage{graphicx}
\usepackage{amsmath,amsthm,mathtools}
\usepackage{kpfonts}
\usepackage[ruled]{algorithm}
\usepackage{algorithmicx}
\usepackage{algpseudocode}
\usepackage{soul, xcolor, todonotes, easyReview}
\usepackage{helvet}         
\usepackage{courier}        
\usepackage{type1cm}        
\usepackage{makeidx}         
\usepackage{graphicx}        
\usepackage{multicol}        
\usepackage[bottom]{footmisc}

\usepackage{pgfplots} \pgfplotsset{compat=1.18}

\newcommand {\mymarginpar}[1]{\marginpar{#1}}
\renewcommand {\marginpar}[1]{} 

\def\_{\rule{.3em}{.15ex}}      

\newcommand{\ls}[1]
   {\dimen0=\fontdimen6\the\font
    \lineskip=#1\dimen0
    \advance\lineskip.5\fontdimen5\the\font
    \advance\lineskip-\dimen0
    \lineskiplimit=.9\lineskip
    \baselineskip=\lineskip
    \advance\baselineskip\dimen0
    \normallineskip\lineskip
    \normallineskiplimit\lineskiplimit
    \normalbaselineskip\baselineskip
    \ignorespaces
   }

\newcommand {\bearn}{\begin{eqnarray*}}
\newcommand {\eearn}{\end{eqnarray*}}
\newcommand {\barr}{\begin{array}}
\newcommand {\earr}{\end{array}}

\newcommand {\N}{{\cal N}}

\newcommand\var[1]{\sigma_{#1}}

\def\defeq{\stackrel{\rm\scriptstyle def}{=}}

\newtheorem{definition}{Definition}
\newtheorem{property}[definition]{Property}
\newtheorem{proposition}[definition]{Proposition}
\newtheorem{lemma}[definition]{Lemma}
\newtheorem{theorem}[definition]{Theorem}
\newtheorem{corollary}[definition]{Corollary}
\newtheorem{example}[definition]{Example}
\newtheorem{remark}[definition]{Remark}
\newtheorem{conjecture}[definition]{Conjecture}
\newtheorem{assumption}[definition]{Assumption}
\newtheorem{approximation}[definition]{Approximation}

\newcommand {\benum} {\begin{enumerate}}
\newcommand {\eenum} {\end{enumerate}}

\newcommand {\bdesc} {\begin{description}}
\newcommand {\edesc} {\end{description}}

\newcommand {\bfig}[2] {\begin{figure}[htbp]
                        \centerline {
                         \epsfig{figure={#1},clip=,width={#2}}}}
\newcommand {\brotatefig}[2] {\begin{figure}[htbp]
                        \centerline {
                         \epsfig{figure={#1},clip=,angle=-90,width={#2}}}}

\newcommand {\bfigfirst}[2] {\begin{figure}[h]
                        \centerline {
                        \setlength{\epsfxsize}{#2}
                        \epsffile{#1}}}
\newcommand {\efig}[2]{ \caption{#2}
                        \label{fig:#1}
                        \end{figure}
                        \mymarginpar{fig:#1}}
\newcommand {\erotatefig}[2]{ \caption{#2}
                        \label{fig:#1}
                        \end{figure}
                        \mymarginpar{fig:#1}}
\newcommand {\rfig}[1]{Figure \ref{fig:#1}}

\newcommand {\btab}[1]{
                       \begin{table}
                       \centering
                       \begin{tabular}{#1}}
\newcommand {\etab}[3] {
                       \end{tabular}
                       \caption[#3]{#2}
                       \label{tab:#1}
                       \end{table}
                       \mymarginpar{tab:#1}
                       \vspace{.1in}}

\newcommand {\btabular}[1]{\begin{center}
                       \begin{tabular}{#1}}
\newcommand {\etabular}{\end{tabular}
                       \end{center}}

\newcommand {\bdefin}[1]{\begin{definition}
                      \mymarginpar{def:#1}
                      \label{def:#1} }
\newcommand {\edefin}       {\end{definition}}

\newcommand {\bassum}[1]{\begin{assumption}
                      \mymarginpar{ass:#1}
                      \label{ass:#1} }
\newcommand {\eassum}       {\end{assumption}}
\newcommand {\rassum}[1]{Assumption \ref{ass:#1}}
\newcommand {\bapprox}[1]{\begin{approximation}
		\mymarginpar{app:#1}
		\label{app:#1} }
	\newcommand {\eapprox}       {\end{approximation}}

\newcommand {\bpro}[1]{\begin{property}
                      \mymarginpar{pro:#1}
                      \label{pro:#1} }
\newcommand {\epro}   {\end{property}}

\newcommand {\bprop}[1]{\begin{proposition}
                      \mymarginpar{prop:#1}
                      \label{prop:#1} }
\newcommand {\eprop}       {\end{proposition}}
\newcommand {\rprop}[1]{Proposition \ref{prop:#1}}

\newcommand {\blem}[1]{\begin{lemma}
                      \mymarginpar{lem:#1}
                      \label{lem:#1} }
\newcommand {\elem}   {\end{lemma}}
\newcommand {\rlem}[1]{Lemma \ref{lem:#1}}

\newcommand {\bthe}[1]{\begin{theorem}
                      \mymarginpar{the:#1}
                      \label{the:#1} }
\newcommand {\ethe}   {\end{theorem}}
\newcommand {\rthe}[1]{Theorem \ref{the:#1}}

\newcommand {\bcor}[1]{\begin{corollary}
                      \mymarginpar{cor:#1}
                      \label{cor:#1} }
\newcommand {\ecor}   {\end{corollary}}

\newcommand {\bax}[1]{\begin{axiom}
                      \mymarginpar{ax:#1}
                      \label{ax:#1} }
\newcommand {\eax}       {\end{axiom}}

\newcommand {\bconj}[1]{\begin{conjecture}
                      \mymarginpar{conj:#1}
                      \label{conj:#1} }
\newcommand {\econj}       {\end{conjecture}}

\newcommand {\bex}[2]{\vspace{.1in}
                      \begin{example}
                      \mymarginpar{ex:#1}
                       {\bf #2}
                      \label{ex:#1} \em}
\newcommand {\eex}       {\end{example} \vspace{.3cm} }

\newcommand {\brem}[1]{\begin{remark}
                      \mymarginpar{rem:#1}
                      \label{rem:#1} }
\newcommand {\erem}   {\end{remark}}

\newcommand {\beq}[1]{\mymarginpar{eq:#1}
                      \begin{equation}
                      \label{eq:#1} }

\newcommand {\beqno}[1]{\mymarginpar{eq:#1}
                      \begin{eqnarray}
                      \nonumber}

\newcommand {\eeq}       {\end{equation}}
\newcommand {\eeqno}       { && \end{eqnarray}}
\newcommand {\req}[1]{(\ref{eq:#1})}

\newcommand {\bear}[1]{\mymarginpar{eq:#1}
                       \begin{eqnarray}
                       \label{eq:#1} }

\newcommand {\bearno}[1]{\mymarginpar{eq:#1}
                       \begin{eqnarray}
                       \nonumber}

\newcommand {\eear}{\end{eqnarray}}
\newcommand {\eearno}{\end{eqnarray}}
\newcommand {\bsel}{\left \{ \begin{array}{cl}}
\newcommand {\esel}{\end{array} \right.}

\newcommand {\bmat}[1]{\left [ \begin{array}{#1}}
\newcommand {\emat}{\end{array} \right ]}

\def\R{I\kern-0.30em R}
\def\N{I\kern-0.30em N}
\def\P{I\kern-0.30em P}

\newcommand {\bxfig}[2] {\begin{figure}[htbp]
                        \centerline {
                         \includegraphics[width=#2]{#1}}}
\newcommand {\brotatexfig}[2] {\begin{figure}[htbp]
                        \centerline {
                         \includegraphics[width=#2,angle=90]{#1}}}
\DeclareGraphicsExtensions{.pdf,.jpg,.png}

\def\ex{{\bf\sf E}}
\def\P{{\bf\sf P}}
\def\pr{\mbox{Pr}}

\def\argsup{\mathop{\rm argsup}}

\def\etal{{\em et al.}\ }
\def\ie{{\em i.e.}\ }

\newcommand{\rcgi}[2]{{#1}_{#2}^{\leftarrow}}
\newcommand{\rcui}[2]{{#1}_{#2}^{\rightarrow}}
\newcommand{\pL}[1]{p_{#1,\mathrm{L}}}
\newcommand{\pT}[1]{p_{#1,\mathrm{T}}}

\newcommand{\pI}[1]{p_{#1,\mathrm{I}}}
\newcommand{\pF}[1]{p_{#1,\mathrm{F}}}

\newcommand{\Iperiod}[2]{{#1}_{#2,\mathrm{I}}}
\newcommand{\Fperiod}[2]{{#1}_{#2,\mathrm{F}}}

\renewenvironment{thebibliography}[1]{
	\begin{oldthebibliography}{#1}
		\setlength{\itemsep}{1pt plus 0.5pt minus 0.5pt} 
		\setlength{\parskip}{0pt}
	}{
	\end{oldthebibliography}
}

\begin{document}

\bstctlcite{IEEEexample:BSTcontrol}	

\title{A Resource Allocation Game and its Equilibrium Strategies
	\footnote{This is the unabridged, extended version of a manuscript submitted to IEEE TCNS. It contains complete mathematical proofs and a first-order mean-field analysis that were omitted from the journal version due to space constraints.}}
\author{Duan-Shin Lee~\IEEEmembership{Senior Member,~IEEE}, 
	\IEEEcompsocitemizethanks{\IEEEcompsocthanksitem 
		D.-S. Lee 
		is with the Department of Computer Science, 
		National Tsing Hua University.
		(Email: lds@cs.nthu.edu.tw)}
	\thanks{This research was supported in part by the Ministry of Science and Technology,
		Taiwan, R.O.C., under Contract 114-2221-E-007-080.}} 
\IEEEtitleabstractindextext{
	\begin{abstract}
	In this paper\footnote{This is the unabridged, extended version of a manuscript submitted to IEEE TCNS. It contains complete mathematical proofs and a first-order mean-field analysis that were omitted from the journal version due to space constraints.} we propose a Bayesian game to allocate resources.
	In this game, there are $c$ units of resources to be
	allocated to $n$ players.  Agent $i$ has a demand of $V_i$ units of resources
	and takes action $X_i$ according to a strategy function $s_i$, i.e. $X_i=s_i(V_i)$.
	Payoffs are setup such that player $i$ is contented with no more than $V_i$
	units of resources.  We assume that resources are granted to the players
	on a smallest-request-first and all-or-nothing basis. 
	For this game with two players, we analyze the equilibrium strategy
	functions mathematically within the family of alternating identity-and-flat (AIF)
	functions. We show that Nash equilibrium profiles consist of two identity functions,
	two AIF functions with a common switch point, or two AIF functions with one, two
	or three switch points, respectively.
	For an $n$-player game with a large $n$ and a large $c_n$ 
	of order $O(n)$, we present a second-order Gaussian approximation for its
	equilibrium strategy function.  In the Gaussian analysis of large games, we
	propose a construction algorithm.  This construction algorithm begins in searching
	within the family of AIF functions.  If a gradient conflict condition occurs, the game
	enters a chattering regime, in which players
	play a continuous, strictly increasing strategy function that is not an identity nor a flat function.
	Conceptually one can view the chattering regime as if players alternate between
	a slope-one strategy and a flat strategy infinitely fast in order to sustain a high payoff.
	We prove that the construction algorithm always
	obtains a Nash equilibrium and terminates in a finite number of steps.  
	We present several numerical examples for the two player game as well as the Gaussian model.
	\end{abstract}
	}

\maketitle

\IEEEdisplaynontitleabstractindextext
\IEEEpeerreviewmaketitle

\section{Introduction}\label{s:introduction}	

Resource allocation has been an important topic for many disciplines, especially
in engineering.  Typically, as new technologies emerge resource allocation
becomes a recurring research subject for the new technologies.  
In the field of information and communications technology (ICT), there
exists a large body of research on resource allocation for cloud computing
\cite{abid2020challenges}, Internet of Things (IoT)~\cite{li2020review}, and 
networking~\cite{xu2021survey}.  In these studies, resources can be physical
quantities such as monetary benefits, computing power, energy/power, 
storage, bandwidth/channel/spectrum and etc.  
Resources can also be non-physical, such as services, access permission and etc.
Classically, resource allocation is formulated as a single-decision making 
problem \cite{fossati2018fair}.  Resource allocation can be rule based.
Popular rules include the proportional fair and the weighted proportional 
fair \cite{Kelly1998}, the max-min fair allocation \cite{Ogryczak2014}, 
and the $\alpha$-fair allocation \cite{Mo2000}.  There is also a large
body of literature that formulates resource allocation as constrained
nonlinear programs.  Possible objective functions include the Jain's
fairness index \cite{Jain1984}, the variance of allocated resources \cite{Lee2023}
and etc. Most resource allocation algorithms in the literature are centralized.
However, there is a class of distributed algorithms for resource allocation.
We refer the reader to \cite{Deng2021, Shao2022, bar1992distributed}
and the references therein. We remark that in these approaches players are non-autonomous
and passively receive allocated resources.  

Alternatively, game-theoretic models allow players to be autonomous, providing a powerful 
paradigm for \textcolor{black}{distributed and networked control systems where agents must make strategic, 
local decisions based on partial information \cite{marden2015game}. Consequently, game 
theory is increasingly utilized to design and analyze competitive resource allocation and 
coordination mechanisms in networked environments \cite{zhou2025distributed,paccagnan2019nash}.} Game theory 
has been applied to study specific resource allocation problems.
For example, Jie \etal \cite{jie2020game} proposed to formulate 
a resource allocation scheme for a fog-based industrial IoT environment in
a double-stage Stackelberg game and proposed three algorithms to 
achieve Nash equilibria and Stackelberg equilibria. 
Fossati \etal in \cite{fossati2018fair} generalized the concept of 
user satisfaction in resource allocation and considered the set of admissible
solutions for a bankruptcy game.  A bankruptcy game is a coalitional game 
with transferable payoffs \cite{Curiel1987}.  
Both Stackelberg games and coalitional games need complete information.  
In some applications, players may not be willing to reveal their information.
Even if they are, it would imply that signaling channels must be provided in order for players
to exchange information.  This requirement may pose problems in a high-speed
environment.  We refer the reader to Huang \cite{huang2015game} for more
research work on formulation of resource allocation in games with complete
information.

In this paper we formulate resource allocation in Bayesian games with
incomplete information.  
In this game, there are $c$ units of resources to be
allocated to $n$ players.  Player $i$, $i=1, 2, \ldots, n$, has a demand of 
$V_i$ units of resources, where $V_i$ has a probability density function (pdf) of
$f_i$. Player $i$ requests $X_i$ units of resources through a strategy function $s_i$, \ie
$X_i=s_i(V_i)$. We assume that resources are granted to players
from the smallest request first until the resources are exhausted. 
Players are granted on an all-or-nothing basis.  That is, if leftover
resources are not enough to satisfy players, players' requests are denied. 
Player $i$'s payoff is $\min\{X_i, V_i\}-\psi(X_i)$, if request $X_i$ is granted, and
is $-\psi(X_i)$ otherwise.  Function $\psi(x)$ is a convex increasing function 
representing a cost of requesting $x$ units of resources. 
We shall make more assumptions on $\psi$ later in the paper.  
We assume that players do not know the demand values of other players.  
However, they do know the distribution of demand values.
The contributions of this paper are as follows. For a game with only
two players, we mathematically analyze the Nash equilibria of the game.
We consider strategy functions that are concatenation of alternating identity functions
and flat (AIF) functions.  We show that Nash equilibrium profiles within the family
of AIF functions can only consist of two identity functions,
two AIF functions with a common switch point, or two AIF functions with one
and three switch points, respectively. In the second type, strategy functions
of both players, despite that they may be heterogeneous, have one common switch point.  
Alternatively, the AIF strategy function of the player who has a lighter demand has 
one switch point. The equilibrium strategy function of the other player has three switch points.
For $n$-player games with a large $n$ and a large $c_n$ of order $O(n)$, we 
present a second-order Gaussian approximation of the payoff function.  
In the Gaussian analysis of large games, we propose a construction algorithm.  
This construction algorithm begins in searching
within the family of AIF functions.  If a gradient conflict condition occurs, the game
enters a chattering regime, in which players
play a continuous, strictly increasing strategy function that is not an identity nor a flat function.
Conceptually one can view the chattering regime as if players alternate between
a slope-one strategy and a flat strategy infinitely fast in order to sustain a high payoff.
We prove that the construction algorithm always
obtains a Nash equilibrium and terminates in a finite number of steps. 
We present several numerical examples for the two player game as well as the Gaussian model.
Finally we mention that games with incomplete information were used to
model resource allocation.  We refer the reader to \cite{huang2015game}.

The outline of this paper is as follows. In Section \ref{s:2player} we present 
a two-player Bayesian resource allocation game. In Section \ref{ss:CEP} 
we derive the conditional expected payoff of a player, given his demand value
and his action. In Section \ref{ss:ES} we analyze the equilibrium strategy functions 
mathematically.  In Section \ref{s:fluid} we present a first-order mean-field analysis
of large games based on the strong law of large numbers.
In Section \ref{s:CLT} we present a Gaussian analysis of
large games.  We present numerical results in Section \ref{s:numerical}.
Section \ref{s:conclusion} contains the conclusion and final remarks of this paper.

\section{Two-player Resource Allocation Games}\label{s:2player}
In this section we describe a resource request and allocation problem 
and a Bayesian game to solve the problem. We assume that a controller has $c$ units
of divisible resources to be allocated to two players.  Agent $i$ has a demand of
$v_i$ units of resources, where $i=1, 2$.  Suppose that player
$i$ requests $x_i$ units of resources.  Since there are totally $c$ units of resources,
it makes no sense to request more than $c$ units.  Thus, we assume that
$0\le x_i\le c$ for $i=1, 2$. The controller grants requests 
from smaller requests first.  Specifically, the controller grants $\min\{x_1, x_2\}$.
Then, the controller attempts to use the leftover to grant $\max\{x_1, x_2\}$.
If there are enough leftovers, the larger request is served.  Otherwise, the larger 
request is rejected.
If there are ties, we assume that ties are resolved randomly.
We assume that resources that exceed players' demands are useless to the players.
It would be beneficial to player $i$ to make a large request $x_i$ that is close to his/her
demand $v_i$, if the request is granted by the controller.  On the other hand, player $i$ 
may want to make a small request to improve the probability of success,
because small requests are more likely to be granted than large ones.
We remark that smallest job first policies are a commonly used priority scheduling
in many disciplines.  For instance, processes with the smallest
execution times are commonly scheduled first in operating systems \cite{Tanenbaum2008}.  
Shortest queues \cite{Guillemin2014} and customers with smallest 
remaining times \cite{Schrage1966} are commonly served first in queueing systems. 
In healthcare applications, smallest-variance-first rule performs well in 
minimizing a weighted sum of mean patients' waiting times and mean doctor's 
idle times \cite{Kemp2021}. Finally in debt management problems, it is commonly
observed that consumers with multiple debts repay small debts first \cite{Amar2011},
although it may be more wise to repay debts with high interest rates first.

Now we present a Bayesian game \cite{Osborne2004} called resource 
allocation game to address the problem above.  This game has two players.
For each $i$, $i=1, 2$, player $i$ has a demand of $V_i$ units of resources.
We call $V_i$ the value of player $i$.
We assume that $V_1$ and $V_2$ are independent 
non-negative random variables with probability
density functions $f_1$ and $f_2$, respectively, for the two players.
Let $F_i$ be the corresponding cumulative distribution function.  Assume 
that $f_i$ and $F_i$ are known to all players. In this paper we typically use
capital letters to denote random variables, and small letters to denote sample
values of random variables.  In the analysis of two-player games, we assume 
that 
$V_i$ is exponentially distributed with mean $1/\lambda_i$ for some $\lambda_i>0$, $i=1, 2$. 
\textcolor{black}{Note that the exponential assumption is specific to the exact 2-player analysis.}
The action space of player $i$, $i=1, 2$, is the set of real numbers $[0, c]$.
Sample values of demands greater than $c$ are \textcolor{black}{censored} to $c$, since the action
space is $[0, c]$.  Thus, we remark that $f_i$ and $F_i$ are pdf and 
cdf censored/rectified to $c$.
Suppose that the value of a player is $v$ and the player adopts
action $x$.  The payoff to this player is $\min\{x, v\}-\psi(x)$ if the request $x$
is granted, where $\psi$ is a convex increasing function representing a cost of requesting
$x$ units of resources. 
If less than $x$ units of resources are available, the request is
rejected and the payoff is $-\psi(x)$ in this case.  The conditional
expected payoff to this player, given action $x$ and value $v$, is
\beq{condexppayoff}
\min\{x, v\}\Pr(\mbox{request $x$ is granted}\, |\, x, v) -\psi(x).
\eeq
The payoff function encourages players to request a large amount of resources.
However, the smallest request first policy favors small requests and makes
large requests less likely to be granted.  Thus, players have to
consider these two factors when they make their requests.
We assume that $\psi'(c/2)<1$.
Note that $\psi'(x)$ is the marginal cost of requesting $x$. Also note that any
request $x\in [0, c/2]$ is granted with probability one.  Thus, requests in $[0, c/2]$
represent a safe region of requests.
The assumption on $\psi'(c/2)$ implies that the marginal request cost at 
the upper end of the safe region is “moderately small”. This reflects that even 
relatively large safe requests remain cheap compared to the value of the resource; 
the main deterrent to very large requests comes from rejection risk rather 
than extreme direct costs.  We summarize the assumptions 
on $\psi(x)$ in the following assumption.
\bassum{psi}
The cost function $\psi(x)$ is increasing, convex, and twice 
continuously differentiable on $[0, c]$.  In addition, $\psi(0)=0$ and 
$\psi'(c/2)<1$.
\eassum

\subsection{Conditional Expected Payoffs}\label{ss:CEP}
In this section we study equilibrium strategy functions that map the value $v$ of a
player to his action.  Given value $v$, a player chooses to take action $s(v)$
according to a strategy function $s$.  Strategy function $s$ is an equilibrium
strategy, if any player adopting $s$ implies that $s$ is also a best response for 
the other player.  In view of the conditional expected payoff in \req{condexppayoff},
we assume that $s$ satisfies the following assumption.
\bassum{s}
The strategy function $s$ is non-decreasing on $[0, c]$ and $0\le s(v)\le v$. In addition,
$s(0)=0$.
\eassum
Non-decreasing strategy functions imply that 
one should not request less, if one has
a larger demand value.  From the definition of payoffs, clearly there is no incentives to
over request.  Thus, it is reasonable to assume that $s(v)\le v$ for all $v\ge 0$.
The condition that $s(0)=0$ is based on an assumption that players are individually
rational.  That is, one should not request anything, if one has no demand.
We remark that $s$ may not possess an inverse function,
since it may not be strictly increasing.  In this paper we shall use two generalized
inverse functions, called right-continuous lower inverse functions and right-continuous
upper inverse functions.
Specifically, define the right-continuous lower inverse function of $s$ as
\beq{rcgif}
{\rcgi{s}{}}(x)\defeq \inf\{v\in R: s(v)\ge x\}
\eeq
and the right-continuous upper inverse function as
\beq{rguif}
{\rcui{s}{}}(x)\defeq \sup\{v\in R: s(v)\le x\}.
\eeq

We begin to analyze the conditional probability that 
request $x_i$ made by player $i$ is granted,
given that the value and action of this player are $v_i$ and $x_i$, respectively.
We adopt the usual convention that $-i$ denotes the player other than $i$.
Let $X_{-i}$ denote the action adopted by player $-i$, who
adopts strategy function $s_{-i}$.
That is, player $-i$ takes action $X_{-i}$, where 
\beq{XV}
X_{-i}=s_{-i}(V_{-i}),
\eeq
if player $-i$'s value is $V_{-i}$. The conditional probability that player $i$'s request
$x_i$ is granted, given request $x_i$ and value $v_i$, is one if $x_i\le c/2$.
If $x_i> c/2$, request $x_i$ is granted if $X_{-i}>x_i$ or $x_i\le c-X_{-i}$.  
In the first event, player $i$ has a higher priority. In the second event, player $i$ has
a lower priority, but there are 
enough leftover resources to be granted to player $i$. Note that these two
events are disjoint. \textcolor{black}{If $X_{-i}=x_i$, there is a tie and probability of 
	grant is $1/2$.} Thus, for $x_i> c/2$,
the conditional probability that request $x_i$ is successfully granted, given $x_i$ and $v_i$, is
\textcolor{black}{
\begin{align}
	&\pr(\mbox{request $x_i$ is granted} | x_i, v_i) \nonumber\\
	&=\pr(X_{-i}>x_i)+\pr(X_{-i}\le c-x_i)+\frac{1}{2}\pr(X_{-i}=x_i)\nonumber\\
	&=1-\frac{1}{2}F_{-i}\left(\rcui{s}{-i}(x_i)\right)-
	\frac{1}{2}F_{-i}\left(\rcgi{s}{-i}(x_i)^-\right) 
	+F_{-i}\left(\rcui{s}{-i}(c-x_i)\right),\label{probgrant1}
\end{align}
where $\rcgi{s}{-i}(x_i)^-$ is the left limit of the generalized inverse.
If $x_i\in (c/2,c]$, the preceding probability is equal to one if $x_i\le c/2$. 
Denote the conditional probability of grant
by $r_i(x_i; s_{-i})$, \ie
\begin{align}
&r_{i}(x_i; s_{-i})=\nonumber\\
&\left\{\begin{array}{ll}
	1, & 0\le x_i\le c/2, \\
	1-\frac{1}{2}F_{-i}\left(\rcui{s}{-i}(x_i)\right)-
	\frac{1}{2}F_{-i}\left(\rcgi{s}{-i}(x_i)^-\right) & \\
	\quad +F_{-i}\left(\rcui{s}{-i}(c-x_i)\right),&
	c/2< x_i\le c,\end{array}\right.\label{probgrant2}
\end{align}
where $v_i$ is player $i$'s demand value.}
From \req{condexppayoff}, player $i$'s conditional expected payoff, given $x_i$ and $v_i$, is
\beq{cp}
u_i(x_i, v_i)=\min\{x_i, v_i\}\textcolor{black}{r_{i}(x_i; s_{-i})}-\psi(x_i).
\eeq
Note that we suppress the dependency of $s_{-i}$ in notation $u_i(x_i, v_i)$ to improve readability.
Define unconstrained expected payoff function
\beq{unconstrained-payoff}
\textcolor{black}{p_{i}(x_i; s_{-i})}=x_i \textcolor{black}{r_{i}(x_i; s_{-i})}-\psi(x_i).
\eeq
Given player $-i$'s strategy function $s_{-i}$, the best response of player $i$ is
\beq{br}
\textcolor{black}{b_i(v_i; s_{-i})}=\argsup_{0\le x\le c} u_i(x_i, v_i).
\eeq

\subsection{Equilibrium Strategies}\label{ss:ES}

In this section we explore equilibrium strategy functions of the two-player
resource allocation games. First, the probability of success $r_i$ 
is non-increasing.  We report this claim in the
following proposition. We present the proof of this proposition in Appendix A.
\bprop{decp}
For any non-decreasing function $s_{-i}$, \textcolor{black}{$r_{i}(x_i; s_{-i})$}
defined in \eqref{probgrant2} is non-increasing in $x_i$.  
\eprop
Due to \rprop{decp}, we have
\begin{align}
	\sup_{0\le x_i\le c} u_i(x_i, v_i) &= \max\left\{\sup_{0\le x_i\le v_i} u_i(x_i, v_i),
	\sup_{v_i\le x_i\le c} u_i(x_i, v_i)\right\} \nonumber\\
	&= \max\left\{\sup_{0\le x_i\le v_i} p_i(x_i; s_{-i}),p_i(v_i; s_{-i})\right\} \nonumber\\
	&= \sup_{0\le x_i\le v_i}p_i(x_i; s_{-i}).
\end{align}

Thus, the best response is the best strategy within the interval $[0, v_i]$, \ie
\beq{br1}
\textcolor{black}{b_i(v_i; s_{-i})}=\argsup_{0\le x_i\le v_i} \left\{x_i 
\textcolor{black}{r_{i}(x_i; s_{-i})}-\psi(x_i)\right\}.
\eeq
\textcolor{black}{Let $BR_i(s_{-i})$ denote the strategy-valued best response, defined as the set of all strategy 
functions $s_i(\cdot)$ such that $s_i(v_i) = b_i(v_i; s_{-i})$ for all $v_i$. The Nash equilibrium 
is thus formally characterized by the coupled fixed point $s_i \in BR_i(s_{-i})$ for $i=1,2$.}
\begin{remark}\label{record-high-principle}
The best response $\textcolor{black}{b_i(v_i; s_{-i})}$ is determined by identifying the global maximum of the 
unconstrained payoff function $\textcolor{black}{p_i(x; s_{-i})} = x \textcolor{black}{r_i(x; s_{-i})} - \psi(x)$ on the interval $[0, v_i]$. 
We call the points where these global maxima occur the ``record highs" of the payoff function.
As the demand value $v_i$ increases from $0$ to $c$, the behavior of $b_i(v_i; s_{-i})$ is 
governed by the evolution of these record highs:	
\begin{itemize}
	\item Aggressive Phase (Identity): If \textcolor{black}{$p_i(x; s_{-i})$} is strictly increasing at $x = v_i$, 
	the record high on the current interval $[0, v_i]$ occurs at the right boundary. 
	In this case, $\textcolor{black}{b_i(v_i; s_{-i})} = v_i$, and the player adopts an identity strategy, 
	requesting their full demand.
	\item Conservative Phase (Flat): As soon as $v_i$ passes a local maximum 
	(say, at point $t$), the payoff function begins to decline. For values of 
	$v_i$ slightly larger than $t$, the global maximum on $[0, v_i]$ remains 
	stuck at $t$. Consequently, the best response becomes $\textcolor{black}{b_i(v_i; s_{-i})} = t$. This 
	creates a flat interval, where the player caps their request to avoid the 
	excessive risk or cost associated with their actual demand.
	\item Recovery: A flat interval only terminates if $v_i$ reaches a point where 
	the payoff $\textcolor{black}{p_i(x; s_{-i})}$ recovers and exceeds the previous record high $\textcolor{black}{p_i(t; s_{-i})}$, at 
	which point the player may return to an identity strategy.
\end{itemize}
This principle is fundamental to our analysis, as it dictates the alternating 
structure of identity and flat segments in both two-player games and the subsequent
Gaussian limit for $n$ players.
\end{remark}
\noindent \textcolor{black}{We emphasize that we do not restrict the players' strategy spaces a priori to any specific class of functions. Rather, the emergence of alternating identity and flat segments—and, as we will show in Section \ref{s:CLT}, the potential transition into a chattering regime—is a strict mathematical consequence of optimizing the unconstrained payoff function under the smallest-request-first policy.}

Because of the record-high principle, we are particularly interested
in a class of functions we call alternating identity and flat (AIF) functions.
We now formally define this class of functions.
Fix an integer $m\ge 1$. Let $\tau_0=0$ and $\tau_0<\tau_1<\cdots<\tau_m<\tau_{m+1}=c$.
We call $\tau_1, \tau_2, \ldots, \tau_m$ 
{\em switch points}. The interval $[0, c]$ is partitioned into $m+1$
contiguous segments, \ie
\[
[\tau_0, \tau_1), [\tau_1, \tau_2), \ldots, [\tau_m, \tau_{m+1}].
\]
A function $s$ is called an alternating identity-and-flat function with $m$ switch points 
(abbreviated as AIF-$m$), if $s(x)=x$ for 
$x\in [\tau_j, \tau_{j+1})$ where $j$ is even. We call these intervals identity
intervals. \textcolor{black}{For odd valued $j$,
$s(x)=s(\tau_{j}^-)$ for $x\in [\tau_{j},\tau_{j+1})$}, where $s(x^-)$ denotes the left 
limit of $s$ at $x$. We call these intervals flat intervals.

An AIF function is nondecreasing and is characterized by its switch points.
Behaviorally speaking, a player who adopts an AIF strategy is aggressive
if his/her value is in an identity interval, and is conservative or holds back if
his/her value is in a flat interval.  
In this paper we are interested in equilibrium profiles formed with two AIF functions
with $m_1$ and $m_2$ switch points, respectively. According to the definition above,
an identity function on $[0,c]$ does not have switch points.  However, 
we call the identity function $s(x)=x$ for $x\in[0, c]$ an AIF-0 function.
\textcolor{black}{An graphical illustration of an AIF-1 and an AIF-3 function is 
shown in Fig. \ref{fig:aif_strategies}.}

\begin{figure}[!t]
	\centering
	\begin{tikzpicture}
		\begin{axis}[
			width=0.98\columnwidth,
			height=5cm, 
			axis line style={black, thin},
			tick align=inside,
			tick style={thin, black},
			grid=both,
			grid style={line width=0.4pt, draw=gray!30},
			xlabel={demand ($v$)},
			ylabel={action ($s(v)$)},
			label style={font=\footnotesize},
			tick label style={font=\footnotesize},
			xmin=0, xmax=2.0,
			ymin=0, ymax=2.0,
			legend pos=north west,
			legend style={font=\scriptsize, fill=white, fill opacity=0.8, draw opacity=1}
			]
			
			\addplot[
			color=red!80!black, 
			dashed, 
			line width=0.8pt,
			domain=0:2.0, 
			samples=2
			] {x};
			\addlegendentry{$s(v)=v$}
			
			\addplot[
			color=blue!80!black, 
			line width=1.5pt
			] coordinates {
				(0, 0)
				(1.17062, 1.17062) 
				(2.0, 1.17062)
			};
			\addlegendentry{AIF-1}
			
			\addplot[
			color=green!60!black, 
			line width=1.5pt
			] coordinates {
				(0, 0)
				(1.17062, 1.17062) 
				(1.23076, 1.17062) 
				(1.23076, 1.23076) 
				(1.44239, 1.44239) 
				(2.0, 1.44239)
			};
			\addlegendentry{AIF-3}
			
		\end{axis}
	\end{tikzpicture}
	\caption{Structure of the AIF-1 and AIF-3 equilibrium 
		strategy functions. Both AIF functions cap at 1.17062. The AIF-3 strategy overlaps the 
		AIF-1 strategy until the second switch point (1.23076), where it jumps back to the 
		identity function before capping again at the third switch point (1.44239).}
	\label{fig:aif_strategies}
\end{figure}

We now derive expected payoffs for player $i$ taking action $x$ and having value $v$.
Since the probability of acceptance $r_i(x; s_{-i})$ is one for $x\in [0, c/2]$, 
we assume that $x\in (c/2, c]$. Because of the record-high principle (Remark
\ref{record-high-principle}), we assume that $v\ge x$.
First, suppose that player $-i$ takes the identity strategy in the neighborhood of $x$.
Denote the expected payoff in this case by $\pI{i}(x)$.
\textcolor{black}{Clearly $\rcui{s}{-i}(x)=\rcgi{s}{-i}(x)^-=x$. In addition, $c-x<c/2$ and
$\rcui{s}{-i}(c-x)=c-x$.} Thus, we have
\begin{align}
\pI{i}(x)&=x\left(1-F_{-i}(x)+F_{-i}(c-x)\right)-\psi(x) \nonumber\\
&= x\left(1+e^{-\lambda_{-i}x}-e^{-\lambda_{-i}(c-x)}
\right)-\psi(x).\label{pI}	
\end{align}
\textcolor{black}{Next, consider the case that player $-i$'s strategy function has a flat interval
$[\tau_j, \tau_{j+1}]$, where $\tau_{j+1}<c$. If $x\in (\tau_j, \tau_{j+1}]$, $\rcui{s}{-i}(x)=\tau_{j+1}$,
$\rcgi{s}{-i}(x)^-=\tau_{j+1}$, and $\rcui{s}{-i}(c-x)=c-x$.
If $x=\tau_j$, $\rcui{s}{-i}(x)=\tau_{j+1}$,
$\rcgi{s}{-i}(x)^-=\tau_j$, and $\rcui{s}{-i}(c-x)=c-x$. 
Denote the expected payoff of player $i$ by
$\pF{i}(x, \tau_j, \tau_{j+1})$. We define
\begin{equation}
\pF{i}(x, \tau_j, \tau_{j+1})=\begin{cases}
\pT{i}(\tau_j, \tau_{j+1}) & \mbox{if $x=\tau_j$,} \\
\pL{i}(x, \tau_{j+1}) & \mbox{if $\tau_j<x\le \tau_{j+1}$,}\\
\pI{i}(x), & \text{if}\ x\in(\tau_j-\epsilon, \tau_j) \\
 & \text{or}\ 
		x\in(\tau_{j+1}, \tau_{j+1}+\epsilon).\end{cases}\label{pF}
\end{equation}
For $x=\tau_j$, define
\begin{align}
&\pT{i}(\tau_j, \tau_{j+1})\nonumber\\
&=\tau_j\left(1-\frac{1}{2}F_{-i}(\tau_{j+1})-\frac{1}{2}F_{-i}(\tau_j)
		+F_{-i}(c-\tau_j)\right)-\psi(\tau_j), \nonumber\\
&= \tau_j\left(1+\frac{1}{2}e^{-\lambda_{-i}\tau_{j+1}}+
\frac{1}{2}e^{-\lambda_{-i}\tau_j}-e^{-\lambda_{-i}(c-\tau_j)}
\right)-\psi(\tau_j).\label{pT}
\end{align}
Similarly for $x\in (\tau_j, \tau_{j+1}]$ define
\begin{align}
\pL{i}(x, \tau_{j+1})&= x\left(1-F_{-i}(\tau_{j+1})+F_{-i}(c-x)\right)-\psi(x)\nonumber\\
&= x\left(1+e^{-\lambda_{-i}\tau_{j+1}}-e^{-\lambda_{-i}(c-x)}\right)
-\psi(x).\label{pL}
\end{align}
Note that in the special case where $\tau_{j+1}=c$, $F_{-i}(c)=1$ and
\begin{align}
\pT{i}(\tau_j, c)&= \tau_j\left(1+\frac{1}{2}e^{-\lambda_{-i}\tau_j}-e^{-\lambda_{-i}(c-\tau_j)}
\right)-\psi(\tau_j),\label{pTsc}\\
\pL{i}(x, c) &= x\left(1-e^{-\lambda_{-i}(c-x)}\right)
-\psi(x),\quad \tau_j<x\le\tau_{j+1}.\label{pLsc}
\end{align}
}

Now we begin analyzing the equilibrium strategy functions for the two-player game.
Let $v\in [0, c/2]$. Since $r_i(x; s_{-i})=1$ for $0\le x\le v$, the payoff is $x-\psi(x)$, and is
strictly concave.  Since $1-\psi'(x)\ge 1-\psi'(c/2)>0$, the maximum of $p_i(x; s_{-i})$
on interval $[0, v]$ occurs at the right boundary $v$.  Thus, the identity function is a dominant
strategy function no matter what opponent's strategy $s_{-i}$ is for $v\in [0, c/2]$. 
Next we consider $v\in (c/2, c]$. 
We present a construction algorithm for a Nash equilibrium profile $(s_1, s_2)$.
The following lemma contains properties that are useful to analyze equilibrium strategy functions.  
\blem{lem1}
\begin{enumerate}
	\item Function $\pI{i}$ is strictly concave on $[c/2, c]$ for any $\lambda_{-i}>0$.
	\item \textcolor{black}{Equation
		\beq{aux}
		e^{-y}(1-y)-y-\psi'(y/\lambda_{-i})=0
		\eeq
		has a unique root in $[0,\infty)$.  Denote this root by $\theta_{-i}$. 
		If $\lambda_{-i} c< \theta_{-i}$, 
		\beq{p_L'=0}
		\pI{i}'(v)=0
		\eeq 
		has no root in $[c/2, c]$. If $\lambda_{-i} c= \theta_{-i}$, \req{p_L'=0} has a unique root at $c$.
		In both cases, function $\pI{i}(v)$ is strictly 
		increasing on $[c/2, c]$. If $\lambda_{-i} c> \theta_{-i}$, \req{p_L'=0} has a unique root
		in $[c/2, c)$.  In this case, denote this root by $v_{i}^*$.
	\item Function $\pL{i}(x,c)$ is strictly concave in $x$.
	\item Let $\phi_{-i}$ be the unique root of the equation
	\beq{aux2}
	e^{-y}-y-\psi'(y/\lambda_{-i})=0.
	\eeq
	For $\theta_{-i}<\lambda_{-i}c\le \phi_{-i}$, $\pL{i}'(c,c)\ge 0$ and $\pL{i}(x,c)$ is
	strictly increasing in $x$ and attains its maximum at $x=c$. If $\lambda_{-i}c> \phi_{-i}$,
	then $\pL{i}'(c,c)< 0$. The maximum of $\pL{i}(x,c)$ occurs at a unique interior
	point in $[c/2, c)$.
	\item For any $\tau_{j+1}\in (c/2, c]$, any $i$, and any $x<\tau_{j+1}$,
	\[
	\pI{i}(x)> \pT{i}(x,\tau_{j+1})>\pL{i}(x, \tau_{j+1}).
	\]}
\end{enumerate}
\elem 
\noindent We present the proof of this lemma in Appendix A.

We first consider the case in which $\lambda_1 c\le \textcolor{black}{\theta_1}$ and $\lambda_2 c\le \textcolor{black}{\theta_2}$.
According to \rlem{lem1}, the expected payoffs of both players are strictly increasing 
on $[c/2, c]$. Thus, the equilibrium
profile is the (AIF-0, AIF-0).  If $\lambda_{-i}c > \textcolor{black}{\theta_{-i}}$ 
and $\lambda_{i}c \le \textcolor{black}{\theta_{i}}$, 
let $v_{i}^*$ denote the interior maximizer of $\pI{i}(x)$, and let \textcolor{black}{$v_{-i}^*=c$}. 
In this case the equilibrium profile is 
an (AIF-1, AIF-1), where the two AIF-1 functions have a common switch point at $v_i^*$.
If $\lambda_{-i} c > \textcolor{black}{\theta_{-i}}$ holds for $i=1, 2$, both
players have an interior maximizer in $[c/2, c)$.  In this case, let
$i$ be the player who has the smaller maximizer, \ie $v_i^*\le v_{-i}^*$. 
\textcolor{black}{Let 
\[
M=\sup_{c/2\le x\le c} \pL{i}(x, c).
\]
Let 
\begin{equation}
\ell =\max\left\{c/2,\inf\{x: \pI{i}(x)=M\}\right\} \label{ell}
\end{equation}
Choose $t_{i,1}\in [\ell, v_i^*]$.} We claim that 
\beq{si}
s_i(v)=\min\{v, t_{i,1}\} 
\eeq
is an equilibrium strategy function for player $i$. We also claim that
for $v$ less than $t_{i,1}$ player $-i$'s equilibrium strategy function is
the identity function.  \textcolor{black}{From Statement 5 of \rlem{lem1}, we claim that
$s_i$ in \req{si} is a restricted dominant strategy function as player $-i$'s strategy
space is restricted to functions sharing this initial identity segment.}
For $v\ge t_{i,1}$, player $-i$'s equilibrium strategy
function is a best response to player $i$'s flat strategy capped at $t_{i,1}$.
Now we construct strategy $s_{-i}$ for player $-i$. Since $x-\psi(x)$ is increasing on
$[0, c/2]$, let $s_{-i}$ be the identity function on this interval.  For 
$v\in [c/2, t_{i,1}]$, player $i$ plays an identity strategy function.  
Thus, player $-i$'s payoff function
is $\pI{-i}(v)$, which is increasing.  Thus, let $s_{-i}$ be the identity function
for $v\in [c/2, t_{i,1})$.  For $v\ge t_{i,1}$, player $i$ plays a flat strategy with flat interval
$[t_{i,1}, c]$. Consider action $x$, where $x\in [0, v]$. Player $-i$'s expected payoff
$p_{-i}(x)$ is equal to $\pI{-i}(x)$ if $x\in [0, t_{i,1})$, is equal to \textcolor{black}{$\pT{-i}(t_{i,1},c)$}
if $x=t_{i,1}$, and is equal to \textcolor{black}{$\pL{-i}(x, c)$ }if $x\in (t_{i,1}, c]$. Since $\pI{-i}(x)$ is
increasing, $\pI{-i}(t_{i,1})$ is a record high. From Statement 5 of
\rlem{lem1}, we let $s_{-i}$ be a flat function with flat interval $[t_{i,1}, t_{-i,2}]$,
where $t_{-i,2}$ is to be determined. If equation 
\beq{2ndsw}
\textcolor{black}{\pL{-i}(v,c)=\pI{-i}(t_{i,1})}
\eeq
with unknown $v$ has no root in $(t_{i,1}, c]$, \textcolor{black}{$\pL{-i}(x, c)<\pI{-i}(t_{i,1})$} 
for all $v\in (t_{i,1}, c]$.
In this case set $t_{-i,2}=c$ and $s_{-i}$ is an AIF-1 function with switch
point $t_{-i,1}=t_{i,1}$, \ie
\beq{s-i-1}
s_{-i}(v)=\min\{v, t_{i,1}\}.
\eeq  
On the other hand, if \req{2ndsw} has a root in 
$(t_{i,1}, c]$, let $t_{-i,2}$ denote the smallest root.  $[t_{i,1}, t_{-i, 2}]$ is
a flat interval of function $s_{-i}$. From Statement 3 of \rlem{lem1}, \textcolor{black}{$\pL{-i}(v,c)$}
is strictly concave. If \textcolor{black}{$\theta_i< \lambda_{i}c\le \phi_i$}, \textcolor{black}{$\pL{-i}'(c,c)\ge 0$}. Set
\beq{s-i-2}
s_{-i}(v)=\begin{cases}
v, & v\in [0, t_{i,1}), \\
	t_{i,1}, & v\in [t_{i,1}, t_{-i,2}], \\
	v, & v\in (t_{-i,2}, c].\end{cases}
\eeq
$s_{-i}$ is an AIF-2 function. On the other hand, if \textcolor{black}{$\lambda_{i} c>\phi_i$}, 
\textcolor{black}{$\pL{-i}(x, c)$} has an interior maximizer in $(t_{-i,2}, c)$ and \textcolor{black}{$\pL{-i}'(c,c)<0$}.
Denote this maximizer by $t_{-i, 3}$.
Function \textcolor{black}{$\pL{-i}(v,c)$} is increasing on $[t_{-i,2},t_{-i,3})$ and is decreasing
on $(t_{-i,3},c]$.  Thus, interval $[t_{-i,2},t_{-i,3})$ is an identity interval
and $[t_{-i,3},c]$ is a flat interval.  It follows that $s_{-i}$ is an AIF-3 function.
In summary, we set
\beq{s-i-3}
s_{-i}(v)=\left\{\begin{array}{ll}
	v, & v\in [0, t_{i,1}), \\
	t_{i,1}, & v\in [t_{i,1}, t_{-i,2}], \\
	v, & v\in (t_{-i,2}, t_{-i,3}), \\
	t_{-i,3}, & v\in [t_{-i,3},c].\end{array}\right.
\eeq
We summarize the construction algorithm in Algorithm \ref{alg1}.

\begin{algorithm}
	\footnotesize  
	\caption{Construction Algorithm for $s_1$ and $s_2$}
	\label{alg1}
	\begin{algorithmic}[1]
		\State Input: $c$, $\lambda_1$ and $\lambda_2$
		\If{$\lambda_1 c\le\theta_1$ and $\lambda_2 c\le \theta_2$}
		\State Set $s_1(v)=s_2(v)=v$ on $[0, c]$. \textcolor{black}{{\bf Terminate}}
		\Else
		\State Let $i$ be the index such that $v_i^*\le v_{-i}^*$
		\State \textcolor{black}{Compute $\ell$ according to \eqref{ell}
		\State Pick a point $t_{i,1}$ from $[\ell,v_i^*]$}
		\State Set $s_{i}(v)=\min\{v, t_{i,1}\}$
		\If{$\sup_{t_{i,1}\le v\le c}\textcolor{black}{\pL{-i}(v, c)}< \pI{-i}(t_{i,1})$}
		\State Set $s_{-i}$ according to \req{s-i-1}
		\Else
		\State Set $t_{-i,2}$ to be the root of \req{2ndsw}
		\If{\textcolor{black}{$\lambda_{-i}c\le \phi_{-i}$}}
		\State Set $s_{-i}$ according to \req{s-i-2}
		\Else
		\State Set $t_{-i,3}$ to be the maximizer of \textcolor{black}{$\pL{-i}(x,c)$}
		\State Set $s_{-i}$ according to \req{s-i-3}
		\EndIf
		\EndIf
		\EndIf
	\end{algorithmic}
\end{algorithm}

One main result of this paper is presented in the following theorem.
\textcolor{black}{Its proof is presented in the appendix.}
\bthe{main}
The strategy functions $s_1$ and $s_2$ constructed by Algorithm \ref{alg1}
form a Nash equilibrium.
\ethe

We remark that in view of steps 8 and 10 of Algorithm \ref{alg1}
an asymmetric game, where $\lambda_1\ne\lambda_2$, may have only symmetric
Nash equilibria. A consequence of \rthe{main} is the following corollary.
\bcor{cor1}
The Nash equilibrium profiles constructed by Algorithm \ref{alg1} can be (AIF-0, AIF-0),
(AIF-1, AIF-1), (AIF-1, AIF-2), (AIF-2, AIF-1), (AIF-1, AIF-3), or (AIF-3, AIF-1).
\ecor

\section{First-order Mean-Field Equilibrium Analysis}\label{s:fluid}

In this section we analyze $n$-player resource allocation games.
We consider a special case where $n$ and $c$ are large and players are
homogeneous, \ie their demands follow a common general pdf $f(x)$.  
In this section we present a first-order 
mean field approximations of this game.  We refer the reader
to \cite{huang2006large,lasry2007mean,jovanovic1988anonymous} for 
the first-order approximation of large games.  These papers lay monumental and 
pioneer foundation in the study of large games. 

In this section and the next section we consider a sequence of games.  
In the $n$-th game, there are $n$ players and
the number of resources is $c_n$.  Specifically, we assume that $c_n/n\to c$ as $n\to\infty$.
Since players are homogeneous, we focus on player 1.  Suppose that $x$ and $v$ are the
request and demand of player 1. Request $s(v)$ is granted if the load
\beq{Ldef}
\hat{L}_n(v) = \frac{1}{n-1} \sum_{j=2}^n s(V_j) \cdot \mathbb{I}(V_j \le v)
\eeq
is less than or equal to $(c_n-x)/(n-1)$, where $\mathbb{I}(A)$ is the indicator function
of event $A$. Let $\mu(v)$ be the theoretical expected request per player 
up to value $v$, \ie
\beq{mudef}
\mu(v) = \ex[s(V) \cdot \mathbb{I}(V \le v)] = \int_0^{v} s(t) 
f(t)\, dt.
\eeq
Fix any $v \in [0, c_n]$. The sequence of random variables 
$\{s(V_j) \cdot \mathbb{I}(V_j \le v), j=2, 3, \ldots, n\}$ are independent and identically
distributed (i.i.d.) with finite expectation.
By Kolmogorov's Strong Law of Large Numbers \cite[Corollary 2, p. 125]{ChowTeicher1997}, 
\[
\hat{L}_n(x) \xrightarrow{a.s.} \mu(v) \quad \text{as } n \to \infty.
\]
In addition, since $\mu(x)$ is continuous and monotonic, and $\hat{L}_n(x)$ is a 
monotonic empirical process, 
by the Glivenko-Cantelli Theorem \cite[Theorem 2, p. 266]{ChowTeicher1997}
\beq{uc-mu}
\sup_{x \in [0, \infty)} | \hat{L}_n(v) - \mu(v) | \xrightarrow{a.s.} 0.
\eeq

Now we analyze the equilibrium strategy function in the first order mean-field analysis.
First, $\psi$ imposes a cost limit to the players.  Since $\psi$ is convex,
the payoff $x-\psi(x)$ is concave and has unique maximizer.  Let $\xi$ be the root
of 
\[
1-\psi'(\xi)=0.
\]
A rational player would not play action $x>\xi$ because of the record high principle.
Thus, a tentative equilibrium strategy function is 
\beq{foma}
s(v)=\min\{v, \xi\}.
\eeq

A request $x$ is accepted in the finite system if $\hat{L}_n(v) \le 
\frac{c_n - s(v)}{n-1} \approx c$.
Due to uniform convergence, the probability of success $r^n(x)$ that player 1's request is accepted
in the $n$-th game converges to a step function
\[
r^\infty(x) = \begin{cases} 1, &  x < \hat\xi, \\ 
	0, &  x \ge \hat\xi. \end{cases}
\]
In the first-order mean-field limit, a player faces a deterministic 
success condition: accept if request $x< \hat\xi$, reject if $x\ge \hat\xi$.
The payoff of player 1 for demand $v$ and request $x$ is
\beq{fluid-payoff}
p(x) = \min(x, v) \cdot \mathbb{I}(x < \hat\xi)-\psi(x).
\eeq
The threshold $\hat\xi$ occurs at point where 
\beq{threshold}
\int_0^{\xi} v f(v)\,dv + \xi(1-F(\xi)) = c.
\eeq
Suppose that players take the strategy function in \req{foma}. If the left hand side
of \req{threshold} is strictly less than the right hand side,
resources are abundant and \req{foma} is the equilibrium strategy function.
On the other hand, if the preceding inequality does not hold, the equilibrium
strategy function is $s(v)=\min\{v, \hat\xi\}$, where $\hat\xi$ is the solution of
\req{threshold}.
We may unite the two cases and state the equilibrium strategy function as
\[
s(v)=\min\{v, \min\{\xi, \hat\xi\}\}.
\]

\section{Gaussian Mean-Field Equilibrium Analysis}\label{s:CLT}

In this section we present a Gaussian approximation of the resource allocation games with
a large number of homogeneous players.  \textcolor{black}{We refer the reader
to \cite{delarue2019master,cardaliaguet2019master,bertucci2019fokker} for 
second-order mean-field approximation of large
games.  These papers lay monumental and pioneer foundation in the study of large games.
The primary motivation for introducing this Gaussian approximation is mathematical tractability. The equilibria computed under this approximation represent the asymptotic limit of the original game; for a large but finite $n$, the Gaussian equilibrium serves as an $\epsilon$-Nash equilibrium where the incentive to unilaterally deviate approaches zero.}

We make the following assumptions for technical reasons.
\bassum{analyticity}
The probability density function $f(v)$ and the cost function $\psi(x)$ are 
piecewise analytic on $[0, c_n]$.  In addition, we assume that
$f(v)$ is non-increasing in $v$.
\eassum
\bassum{pdfshape}
The probability density function $f(v)$ is non-increasing in $v$.
In addition, assume that $v^3 f(v)\to 0$ as $v\to\infty$.
\eassum
Let 
\[
Y_j=s(V_j) \mathbb{I}(V_j \le v).
\]
Let $\mu(v)$ and $\sigma(v)$ denotes the mean and the standard deviation of
$Y_j$, \ie
\begin{align}
\mu(v) &= \int_0^v s(t) f(t)\,dt, \label{mudef}	\\
\sigma(v) &= \sqrt{\int_0^{v} s(t)^2 f(t) \, dt - \mu(v)^2}.\label{sigmadef}
\end{align}
Since $\{Y_j, 2\le j\le n\}$ is a sequence of i.i.d. random variables, the central
limit theorem and 
Berry-Esseen Theorem \cite{ChowTeicher1997} imply that 
\begin{align*}
&\pr(Y_2+\ldots+Y_n\le c_n-s(v)) \\
&=\pr\left(\frac{Y_2+\ldots+Y_n-(n-1) \mu(v)}{\sigma(v)\sqrt{n-1}}\le 
\frac{c_n-s(v)-(n-1)\mu(v)}{\sigma(v)\sqrt{n-1}}\right)\\
&\approx\Phi(w(v)),
\end{align*}
where $\Phi$ is the cumulative distribution function of a standard normal random variable and
\beq{wdef}
w(v)=\frac{c_n-s(v)-(n-1)\mu(v)}{\sigma(v)\sqrt{n-1}}.
\eeq
We present two important properties of the z-score $w(v)$ defined above. These properties
will be used in the proofs of several propositions. Let $\mu(\infty)$ and
$\sigma(\infty)$ denote
\begin{align*}
\mu(\infty) &=\int_0^\infty s(v)f(v)\,dv \\
\sigma(\infty) &= \int_0^\infty s(v)^2 f(v)\,dv-\mu(\infty)^2.
\end{align*}
\blem{z-score}
Assume that $c-\mu(\infty)>0$. Let $s(v)$ be any non-decreasing strategy function 
such that $0 \le s(v) \le v$. Further, assume that there exists $v\in [0, c_n]$ such that
$s(v)$ is of order $O(n)$.
\begin{enumerate}
	\item The numerator of $w(v)$ has a unique root denoted by $\tilde v_n$ 
	and $s(\tilde v_n)\approx n[c-\mu(\infty)]$. 
	\item Suppose that $v>\tilde v_n$ and
	$s(v)/n \approx x_0 > c-\mu(\infty)$. Then, $w(v)$ is negative and is of order 
	$O(\sqrt{n}(c-x_0-\mu(\infty))/\sigma(\infty))$.
	\item Under the same assumption stated in the previous Statement, $w'(v)$ is negative 
	and is of order $O(n^{-1/2}/\sigma(\infty))$.
\end{enumerate}
\elem
\noindent We present the proof of \rlem{z-score} in Appendix B.

The probability of success that request $x$ is granted is $\Phi(w(v))$.
The conditional expected payoff, given strategy $x$ and value $v$, is
\beq{constrained-Gaussian-payoff}
\min\{x, v\}\Phi(w(v))-\psi(x).
\eeq
The unconstrained conditional expected payoff of player 1, given value $v$ 
and strategy function $s$, is
\beq{pdef}
p(v)=s(v) \Phi(w(v))-\psi(s(v)).
\eeq
\begin{remark}
{\bf [No-look-ahead Strategy Construction]} 
The property that $p(x)$ depends on $s(v)$ only through the integrals 
on $[0, v]$ is a direct consequence of the Smallest-Request-First 
policy. This causality allows us to construct the equilibrium strategy 
$s(\cdot)$ progressively from $v=0$ upwards without knowledge of the strategy 
for larger demand values. Crucially, this independence holds regardless of the 
specific regime the game enters for higher values. 
We shall present a construction algorithm (in Algorithm \ref{alg2}) 
for equilibrium strategy functions. We search strategy functions within
the family of AIF function until the demand value reach a certain point,
say $\tau\in [0, c_n]$.  We call $[0, \tau]$ an AIF regime.  If a gradient
conflict occurs, the game may enter a chattering regime (CR). Details on both the 
gradient conflict and CRs will be presented later. 
Because of the smallest-first-policy, the equilibrium strategy in CR does not retroactively
alter the moments $\mu(v)$ and $\sigma(v)$ for 
$v \le \tau$ in the AIF regime. Thus, the construction of AIF strategies in 
the AIF regime is completely decoupled from the subsequent analysis of the
chattering regime (CR).
\end{remark}


Since identity intervals and flat intervals alternate, they appear in pairs
except for the last interval.  In this section we shall index them
in pairs. Let $K$ be the number of pairs that function $s$ has.
If the number of switch points $m$ is even, $K=m/2$.  Otherwise, $K=(m+1)/2$.
Suppose that players 2 to $n$ adopt an AIF strategy function with $2m$ or $2m+1$ switch
points. Note that for $v\in [\tau_j, \tau_{j+1})$,
where $j$ is even, $v$ is in an identity interval. In this case, $s(v)=v$ and
$s^\leftarrow(v)=v$. If $j$ is odd ($j=2k-1$), $v$ is in a flat interval.
In this case, $s(v)=\tau_{2k-1}$ and $s^\leftarrow(v)=\tau_{2k}$. 
We denote $s$ on interval $[0, \tau_j]$ by $s_j$. Specifically,
\beq{s-progression}
s_{j+1}(v)=\begin{cases}
	s_j(v) & \mbox{if $v\le \tau_j$}, \\
	v & \mbox{if $j=2k-2$ and $v\in[\tau_j, \tau_{j+1}]$},\\
	\tau_j & \mbox{if $j=2k-1$ and $v\in[\tau_j, \tau_{j+1}]$}.
\end{cases}
\eeq
For $x$ in the $k$-th, $1\le k\le m$, 
identity interval, \ie $v\in [\tau_{2k-2}, \tau_{2k-1})$, denote $\mu(v)$ by
\begin{align}
	\Iperiod{\mu}{k}(v)&=\int_0^{\tau_{2k-2}} s_{2k-2}(t) f(t)\,dt+
	\int_{\tau_{2k-2}}^v t f(t)\,dt. \label{muidentity}
\end{align}
In the beginning of a flat interval,
say $\tau_{2k-1}$, to determine where it ends, we need function $\mu(v)$ which depends on the
end point of this flat interval. At this point we are yet to know where it ends.
For this reason, we define contingent expectation
\beq{muflat}
\Fperiod{\mu}{k}(v)=\int_0^{\tau_{2k-1}} s_{2k-1}(t) f(t)\,dt +
\tau_{2k-1}\cdot\frac{F(v)-F(\tau_{2k-1})}{2}.
\eeq
The validity of $\Fperiod{\mu}{k}(v)$ is contingent on the condition that
the $k$-th flat interval ends at $v$.  
Then, for $v$ in the $k$-th identity interval, we define
\beq{sigmaident}
\Iperiod{\sigma}{k}(v)=\sqrt{\int_0^{\tau_{2k-2}} s_{2k-2}(t)^2 f(t)\,dt+\int_{\tau_{2k-2}}^v t^2 f(t)\,dt
	-\Iperiod{\mu}{k}(v)^2}.
\eeq
We also define contingent standard deviation for $v$ in the $k$-th flat interval
\begin{align}\label{sigmaflat}
&\Fperiod{\sigma}{k}(v)=\nonumber\\
&\sqrt{\int_0^{\tau_{2k-1}} s_{2k-1}(t)^2 f(t)\,dt
+\tau_{2k-1}^2 \frac{F(v)-F(\tau_{2k-1})}{2}-\Fperiod{\mu}{k}(v)^2}.
\end{align}
We define 
\beq{defwk}
\Iperiod{w}{k}(v) =\frac{c_n-v-(n-1)\Iperiod{\mu}{k}(v)}{\Iperiod{\sigma}{k}(v)\sqrt{n-1}}
\eeq
and the unconstrained expected payoff function
\beq{defpk}
\Iperiod{p}{k}(v) = v \Phi(\Iperiod{w}{k}(v))-\psi(v).
\eeq
Note that $v$ in \req{defwk} and \req{defpk} are in an identity interval.  Thus, $v$ is also
the action that player 1 takes.
Functions $\Fperiod{w}{k}$ and $\Fperiod{p}{k}$ for flat intervals are defined as
\begin{align}
	\Fperiod{w}{k}(v) &= \frac{c_n-\tau_{2k-1}-(n-1)\Fperiod{\mu}{k}(v)}
	{\Fperiod{\sigma}{k}(v)\sqrt{n-1}}\label{defwkF}\\
	\Fperiod{p}{k}(v) &=\tau_{2k-1}\Phi(\Fperiod{w}{k}(v))-\psi(\tau_{2k-1}).\label{defpkF}
\end{align}


We present several properties of $w(x)$ and the probability of success $\Phi(w(x))$.
The following proposition provides a justification for simplifying the best-response 
problem to finding record highs of the unconstrained payoff. 
In the proposition we also present properties of $w(x)$ and $w'(x)$. 
\bprop{decp2}
Suppose that $s$ in \eqref{mudef} and \eqref{sigmadef} is an AIF function.  Then,	
\begin{enumerate}
	\item both $\Iperiod{w}{k}(v)$ and $\Fperiod{w}{k}(v)$ are decreasing on $[0, \infty)$; 
	\item the probability of success $\Phi(\Iperiod{w}{k}(v))$ and $\Phi(\Fperiod{w}{k}(v))$
	are decreasing on $[0, \infty)$.
\end{enumerate}
\eprop
\noindent The following proposition is essential to the construction algorithm 
of equilibrium strategy functions.  
\bprop{negdriftatc}
Let $s$ be a non-decreasing equilibrium strategy function mapping player values to actions.
Under the conditions of \rlem{z-score}, the supremum of the equilibrium action space 
is strictly bounded away from the capacity limit $c_n$. That is, there exists 
an $\epsilon > 0$ such that for all $v \in [0, c_n]$
\[s(v) \le c_n - \epsilon.\]
\eprop
\noindent The proofs of the two propositions above are presented in Appendix B.


Since $\Phi(\Iperiod{w}{1}(v))\approx 1$ for
$v\approx 0$ and $\psi'(0)<1$, initially $\Iperiod{p}{1}'(0^+)>0$.
Thus, we begin with an identity interval. According to \rprop{negdriftatc}, 
$\Iperiod{p}{1}'(v)$ must cease to increase strictly
at some point before $c$.  Let $\tau_1$ be this point. Point $\tau_1$ is a local
maximum.  Function $\Iperiod{p}{1}(v)$ is decreasing after $\tau_1$ and has
a negative derivative in the right neighborhood of $\tau_1$.  
This suggests that function $s$ should
switch to a flatness mode at $\tau_1$.  The argument above applies to all
identity intervals.  At the end of the $k$-th identity interval $\tau_{2k-1}$ let 
\beq{rh}
P_k^* = \Iperiod{p}{k}(\tau_{2k-1}).
\eeq
$P_k^*$ is the record high of the expected unconstrained expected payoff on $[0, \tau_{2k-1}]$.
There are two cases after the end of an identity interval.
\begin{enumerate}
	\item If $\Fperiod{p}{k}'(v)<0$ in the right neighborhood of $\tau_{2k-1}$, this 
	implies that $\Fperiod{p}{k}(v)<P_k^*$ for $x>\tau_{2k-1}$. 
	Thus, payoff function $\Fperiod{p}{k}(v)$ also suggest to
	switch from an identity mode to a flatness mode, which reinforces the implication from
	function $\Iperiod{p}{k}(v)$. Thus, the strategy function begins a flat interval.
	\item If $\Fperiod{p}{k}'(v)>0$ in the right neighborhood of $\tau_{2k-1}$, this 
	implies that $\Fperiod{p}{k}(v)>P_k^*$ for $v>\tau_{2k-1}$. In the right neighborhood
	of $\tau_{2k-1}$, function $\Iperiod{p}{k-1}(v)$ is decreasing
	and suggests a flat interval. This implies that the identity strategy used by
	players 2 to $n$ is too aggressive such that the expected payoff of player 1
	is below $P_k^*$.  On the other hand, if players 2 to $n$ take the flat
	strategy, $\Fperiod{p}{k}(v)$ is increasing in the right neighborhood. This implies
	that the flat strategy is too conservative in the sense of confining player 1's
	expected payoff. In this case, the game enters a chattering regime (CR).	
	There must be an intermediate strategy function in $[\tau_{2k-1}, c_n]$
	such that the expected payoff of player 1 is maintained right at $P_k^*$. 
	Conceptually, the chattering regime represents a state where players cannot sustain 
	a record-high payoff by committing to the global identity strategy ($s(v)=v$) or a 
	flat strategy ($s(v)$ is a constant). Instead, the equilibrium strategy $\eta(v)$ is a `singular arc' 
	constructed by alternating between flat segments (slope 0) and unit-slope segments 
	(slope 1). While these unit-slope segments mimic the growth rate of an identity function, 
	they are localized `advancing' steps that allow the strategy to track the unique path 
	defined by the record-high sustaining equation as follows. Let $p_\eta(v)$ denote the
	unconstrained expected payoff of using strategy $\eta(v)$.  Let $\tau=\tau_{2k-1}$. Then,
	\beq{sustain-rec-high}
	p_\eta(v)=\eta(v)\Phi(w_\eta(v))-\psi(\eta(v))=P_k^*,
	\eeq
	for all $v\ge \tau$, where 
	\begin{align}
	w_\eta(v) &=\frac{c_n-\eta(v)-(n-1)\mu(v)}{\sqrt{n-1}\sigma(v)},\label{weta}\\
	\mu(v) &= \int_0^\tau s(t)f(t)\,dt+\int_\tau^v \eta(t)f(t)\,dt,\label{mueta}\\
	\sigma(v) &= \sqrt{\int_0^\tau s(t)^2f(t)\,dt+\int_\tau^v \eta(t)^2f(t)\,dt
	-\mu(v)^2}.\label{sigmaeta}
	\end{align}
	We call $\eta(v)$ a chattering strategy and this regime a chattering regime (CR). 
	We present more details on chattering regimes in Section \ref{s:CR}.
	We remark that the conceptual view that $\eta$ is a 
	mixture of unit-slope functions and flat functions is similar to chattering control in control theory 
	\cite{Zelikin1994}.

\end{enumerate}

In the $k$-th flat interval, we compute
\beq{Mmax}
M_k^*=\sup_{\tau_{2k-1}\le v\le c_n} \Fperiod{p}{k}(v).
\eeq
There are two cases as follows.
\begin{enumerate}
	\item If $M_k^*< P_k^*$, the $k$-th flat interval ends at $c_n$. The overall strategy
	function is a concatenation of the strategy function obtained so far on $[0, \tau_{2k-1}]$
	and a flat function of value $\tau_{2k-1}$ on $[\tau_{2k-1}, c_n]$.
	\item If $M_k^*> P_k^*$, equation
	\beq{flatends}
	 \Fperiod{p}{k}(v)=P_k^*
	\eeq
	must have at least one root. Let $\tau_{2k}$ be the smallest root.
	$[\tau_{2k-1},\tau_{2k})$ is a flat interval. Note that in this case, 
	$\Fperiod{p}{k}'(\tau_{2k})>0$, since the function $\Fperiod{p}{k}(v)$ 
	intersects with $M_k^*$ from below. This implies that the strategy function 
	should enter an identity interval from $\tau_{2k}$.
	From here there are two sub-cases as follows.
	\begin{enumerate}
		\item If $\Iperiod{p}{k+1}'(\tau_{2k}^+)>0$, this confirms that the game switches to
		an identity mode at $\tau_{2k}$.
		\item If $\Iperiod{p}{k+1}'(\tau_{2k}^+)<0$, this contradicts with the implication
		from $\Fperiod{p}{k}'(\tau_{2k}^+)>0$ that an identity interval should begin
		at $\tau_{2k}$.  Thus, the game enters a CR.  
	\end{enumerate}
\end{enumerate}

\subsection{Chattering Regimes}\label{s:CR}

In this section we present more details on chattering regimes.  First we present a
remark on the entry and exit of a CR.  We then present a technical assumption.

\begin{remark}\label{transition-between-AIF-and-CR}
{\bf [Entry and Exit of CR]} At any switch point, say $\tau_{2k-1}$ or $\tau_{2k}$, we say that
a gradient conflict occurs, if 
\beq{grad-contra}
\begin{cases}
	\Iperiod{p}{k}'(\tau_{2k-1}^+) =0\text{ and } \Iperiod{p}{k}''(\tau_{2k-1}) < 0 \\
	\Fperiod{p}{k}'(\tau_{2k-1}^+) >0,\end{cases}\ \mbox{or}\ 
\begin{cases}
	\Iperiod{p}{k+1}'(\tau_{2k}^+) <0 &\\
	\Fperiod{p}{k}'(\tau_{2k}^+) >0. &
\end{cases}	
\eeq
There are two scenarios that a CR can occur.
At the end of an identity interval, say the $k$-th interval, if
gradients of conditional expected payoffs given identity strategies and flat strategies
respectively form a conflict, a CR is entered.  Behaviorally speaking, a player
extremely aggressively plays an identity strategy before $\tau_{2k-1}$. Since 
$\Iperiod{p}{k}'(\tau_{2k-1}+\epsilon)<0$, $\epsilon>0$, this conditional expected payoff suggests that
he/she play cautiously after $\tau_{2k-1}$. On the other hand, $\Fperiod{p}{k}'(\tau_{2k-1}+
\epsilon)>0$ suggests that the player should not be too conservative. Thus, the
player plays a moderately and cautiously aggressive strategy after $\tau_{2k-1}$.
From a behavioral point of view, we call this CR a tempered chattering regime.
In a second scenario, at the end of a flat interval, say the $k$-th flat interval,
again a gradient conflict may occur, \ie $\Iperiod{p}{k+1}'(\tau_{2k}+\epsilon)<0$ and
$\Fperiod{p}{k}'(\tau_{2k}+\epsilon)>0$. The game enters a CR.
Behaviorally speaking, a player is very conservative before $\tau_{2k}$. 
The gradient conflict indicates that neither extreme aggression nor conservativeness
is a good option.  Thus, the player plays a moderately conservative strategy function.
For this reason, we call this CR a resurgent chattering regime.

From \rprop{negdriftatc}, the game exits a CR and
players take a flat strategy between the exit point and $c_n$.
\end{remark}

\bassum{marginal}
{\bf [Marginal profitability of chattering]} Throughout a chattering regime, the
equilibrium strategy $\eta(v)$ satisfies $\Phi(w(v))>\psi'(\eta(v))$.
\eassum

We first present a proposition that address the solution of \req{sustain-rec-high}.
Its proof is presented in Appendix B.
\bprop{eta-exist-incr}
Under \rassum{psi}, \rassum{analyticity}, \rassum{marginal}, large $n$, and the
gradient conflict \req{grad-contra} at an entry point $\tau$,
there exists a unique strategy $\eta(v)$ 
that satisfies the record-high sustaining equation \req{sustain-rec-high} for all $v$ in the 
CR. Furthermore, $\tau\le \eta(v)\le v$ and $\eta(v)$ is strictly increasing.
\eprop

We shall present a constructive view of the chattering strategy $\eta(v)$ to show its similarity
with chattering control. Before we do that, we make a remark.  In the proof of \rprop{eta-exist-incr}
we shall derive the gradient of $\eta(v)$ at the entry point $\tau$, \ie
\beq{etap2}
\eta'(\tau)=\frac{\Fperiod{p}{k}'(\tau^+)}{\Fperiod{p}{k}'(\tau^+)-\Iperiod{p}{k}'(\tau^+)}.
\eeq
For tempered CRs, $\Iperiod{p}{k}'(\tau^+)=0$.  Thus, $\eta'(\tau)=1$. Strategy functions
smoothly transit from an identity function to $\eta(v)$.  Behaviorally speaking, players do not 
abruptly change their behavior; instead, the strategy smoothly branches off from 
the identity function, gently curbing their aggression as the risk and cost increase. 
For resurgent CRs, an entry occurs at $\tau=\tau_{2k}$ because $\Fperiod{p}{k}(x)$ crosses
level $P_k^*$ from below as stated in \req{flatends}. The level crossing at $\tau=\tau_{2k}$ implies
that $\Fperiod{p}{k}'(\tau^+)$ is strictly positive. Thus, $\eta'(\tau)$ is also strictly positive.
This means that the strategy exhibits discontinuous derivative at $\tau$ as it abruptly shifts 
from a flat slope of 0 to a fractional slope. 

We now present a constructive view. We shall show that $\eta(v)$ is a uniform limit of a sequence of
continuous strategy functions formed with functions with unit slopes and zero slopes.
To begin, partition the chattering regime $[\tau, c_n]$ into $m$ intervals for a given integer $m>0$.
Let $\delta=(c_n-\tau)/m$ be the width of an interval. Let the grid points be $v_j=\tau+j\delta$,
$j=0, 1, \ldots, m$.  For each interval $[v_j,v_{j+1})$, calculate the target slope 
based on the gradient conflict ratio at the start of the interval (as defined in \req{etap2}
in the proof of \rprop{eta-exist-incr}):
\beq{slope-target}
\rho_j = \eta'(v_j) \in [0, 1].
\eeq
We define $\eta_m(v)$ for $v \in [v_j, v_{j+1})$ as the continuous concatenation of a 
segment with zero slope and a segment with unit slope. Specifically,
let the width of the segment with unit (resp. zero) slope be $\Iperiod{\delta}{j} =\rho_j\delta$
(resp. $\Fperiod{\delta}{j}=(1-\rho_j)\delta$). Define
\beq{etam}
	\eta_m(v) = 
	\begin{cases} 
		\eta_m(v_j) & \text{for } v \in [v_j, v_j + \Fperiod{\delta}{j}), \\
		\eta_m(v_j) + (v - (v_j + \Fperiod{\delta}{j})) & \text{for } v \in [v_j + \Fperiod{\delta}{j}, v_{j+1}), 
		\end{cases}
\eeq
with initial condition $\eta_m(\tau)=\tau$. 
Clearly by this construction, $\eta_m(v)$ is a continuous and piece-wise linear function on $[0, c_n]$.
We have the following proposition. Its proof is presented in Appendix B.
\bprop{etam-converge}
The sequence $\{\eta_m(v)\}$ converges uniformly to the solution of \req{sustain-rec-high}.
\eprop

We summarize the discussion above in an algorithm. 
We use variable ${\tt mode}$ to keep track of
the type of intervals we are working on.  Identity and flat intervals are denoted
by ${\tt I}$ and ${\tt F}$, respectively, and the CR is
denoted by ${\tt C}$.
Initially, ${\tt mode}$ is set to ${\tt I}$.
We iterate as the ${\tt mode}$ variable alternates
between ${\tt I}$ and ${\tt F}$. The algorithm terminates either in an ${\tt F}$
mode or an ${\tt C}$ model. We summarize the algorithm in Algorithm \ref{alg2}.
We remark that $\coloneqq$ denotes assignment operations.

\begin{algorithm}
	\footnotesize  
	\caption{Equilibrium Construction Algorithm}
	\label{alg2}
	\begin{algorithmic}[1]
		\State Input: $c_n$, $n$ and $f$
		\State Initialization: Set $k\coloneqq 1$, $\tau_0\coloneqq 0$, 
		${\tt mode} \coloneqq {\tt I}$
		\Loop
		\If{${\tt mode}$ = ${\tt I}$}
		\State Find $t$ for the first time $\Iperiod{p}{k}(v)$ ceases to increase
		\State $\tau_{2k-1}\coloneqq t$
		\State $P_k^*\coloneqq \Iperiod{p}{k}(\tau_{2k-1})$
		\If{$\Fperiod{p}{k}'(\tau_{2k-1}^+)<0$}
		\State ${\tt mode}\coloneqq {\tt F}$
		\Else
		\State ${\tt mode}\coloneqq {\tt C}$
		\EndIf
		\EndIf
		\If{${\tt mode}$ = ${\tt F}$}		
		\State Compute $M_k^*$ from \req{Mmax}
		\If{$M_k^* < P_k^*$}
		\State {\bf Terminate}
		\Else
		\State Solve \req{flatends}
		\State Set the smallest root to $\tau_{2k}$
		\If{$\Iperiod{p}{k+1}'(\tau_{2k}^+)>0$}
		\State ${\tt mode}\coloneqq {\tt I}$
		\Else
		\State ${\tt mode}\coloneqq {\tt C}$
		\EndIf
		\EndIf
		\EndIf
		\If{${\tt mode}$ = ${\tt C}$}
		\State Find a chattering equilibrium strategy $\eta(v)$
		\State {\bf Terminate}
		\EndIf
		\State $k\coloneqq k+1$
		\EndLoop
	\end{algorithmic}
\end{algorithm}

The following theorems contain the main results of this section. 
\bthe{finite}
Algorithm \ref{alg2} terminates in a finite number of iterations.
\ethe

\bthe{global-opt}
The strategy function obtained by Algorithm \ref{alg2} is a Nash equilibrium
of the Gaussian mean-field resource allocation game.
\ethe
\noindent The proofs of \rthe{finite} and \rthe{global-opt} are presented in Appendix B.

\section{Numerical Experiments}\label{s:numerical}

We present numerical studies of the two-player resource allocation games for a few cases.
First, we assume that $c=2$, $\lambda_1=1$, $\lambda_2=2$ and $\psi(x)=0$.
In this case both
$\pI{1}(v)$ and $\pI{2}(v)$ have interior maximum points in $[c/2, c]$.  
Function $\pI{2}(v)$ has a smaller
interior maximum point at $v_2^*=1.17062$. In this case, 
\[
\pF{1}(v) < \pI{1}(v_2^*)
\]
for all $v\in [v_2^*, c]$. Thus, $s_1$ is an AIF-1 function with switch point $v_2^*$.
This function is shown in Fig. \ref{fig:aif_strategies} in blue. 
The payoff functions of the two players are shown in dot-dashed lines 
and solid lines \rfig{case1}. 
The equilibrium strategy functions of the two players are identical. 
Next, we increase $\lambda_2$ to $3$.  Strategy function $s_2$ is of the
form of \req{si} and the switch point is at $v_2^*$. Function $s_{1}$ 
is of the form in \req{s-i-3}. The switch points
of $s_1$ are $1.17062$, $1.23076$ and $1.44239$.
The payoff functions of the two players are shown in \rfig{case2}. 
Function $s_1$ is shown in Fig. \ref{fig:aif_strategies} in green. 
In both examples, $\ell=1.0$.  Thus, all points in interval $[1.0, v_2^*]$ can be the first switch
point of the equilibrium strategy functions.

\textcolor{black}{
We also present an example where $s_1$ is of the form in \req{s-i-2}.  We let $\lambda_2=1/4$.
The switch points of $s_1$ are $1.17062$ and $1.5175$.}

\bfig{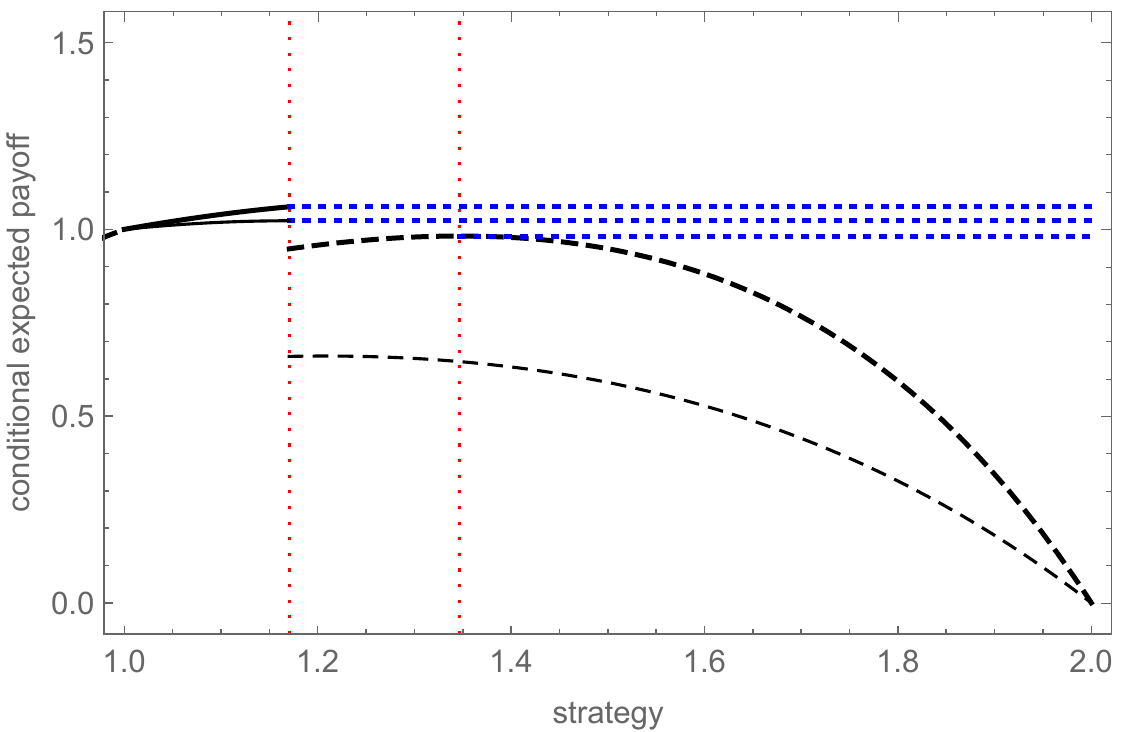}{\columnwidth}
\efig{case1}{Payoff function of player one is drawn in a dot-dashed line and 
payoff function of player two is drawn in a thick solid line. In this case,
$c=2$, $\lambda_1=1$ and $\lambda_2=2$. In this case, $s_1$ and $s_2$ are AIF functions with a 
common switch point at $v_2^*=1.17062$. Functions $\pI{1}$ and $\pI{2}$ are marked with thin and thick
solid curves, respectively. Functions $\pL{1}$ and $\pL{2}$ are marked with thin and thick
dash curves, respectively. This switch point is marked
by a red vertical dashed line. Function values $\pI{1}(v_2^*)$ and $\pI{2}(v_2^*)$ are marked
by horizontal blue dotted lines.  They denote record highs of payoff functions
$p_1(v)$ and $p_2(v)$.}
\bfig{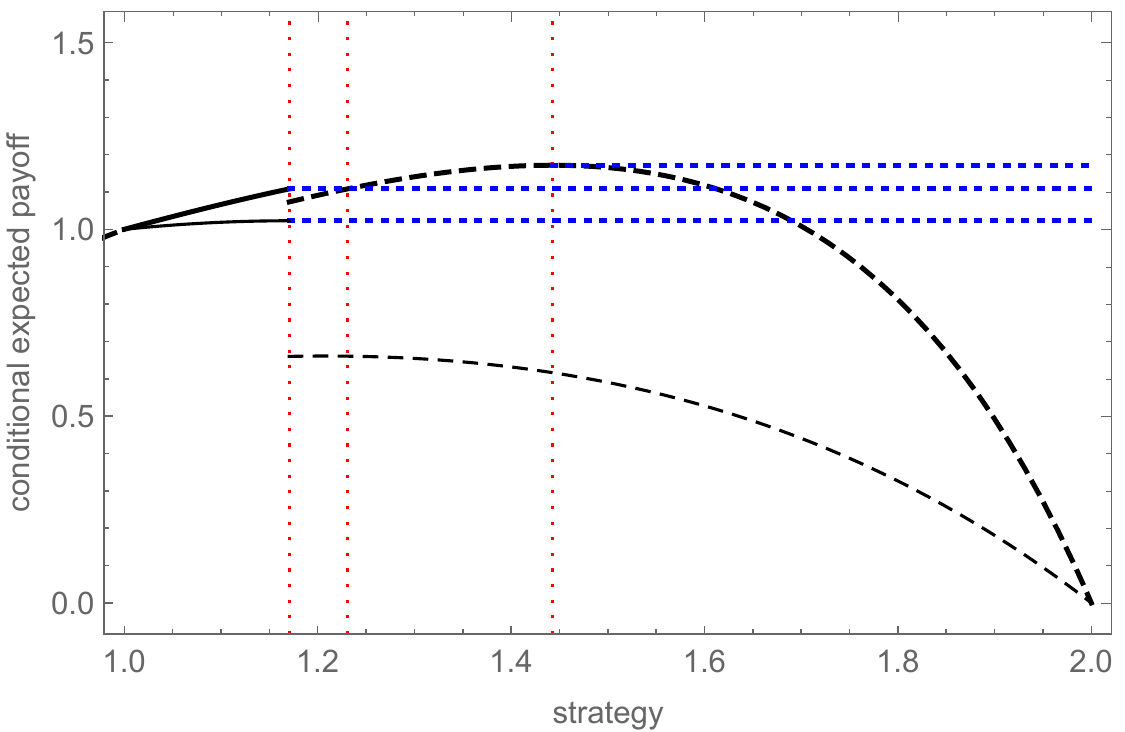}{\columnwidth}
\efig{case2}{Payoff function of player one is drawn in a dot-dashed line and 
	payoff function of player two is drawn in a thick solid line. In this case,
	$c=2$, $\lambda_1=1$ and $\lambda_2=3$. Functions $\pI{1}$ and $\pI{2}$ are marked with thin and thick
solid curves, respectively. Functions $\pL{1}$ and $\pL{2}$ are marked with thin and thick
dash curves, respectively. The switch points of $s_1$ are marked
	by three vertical red dashed lines. Function values $\pI{1}(v_2^*)$, $\pI{2}(v_2^*)$ 
	and $\pF{1}(t_{1,3})$ are marked by horizontal blue dotted lines.}
	
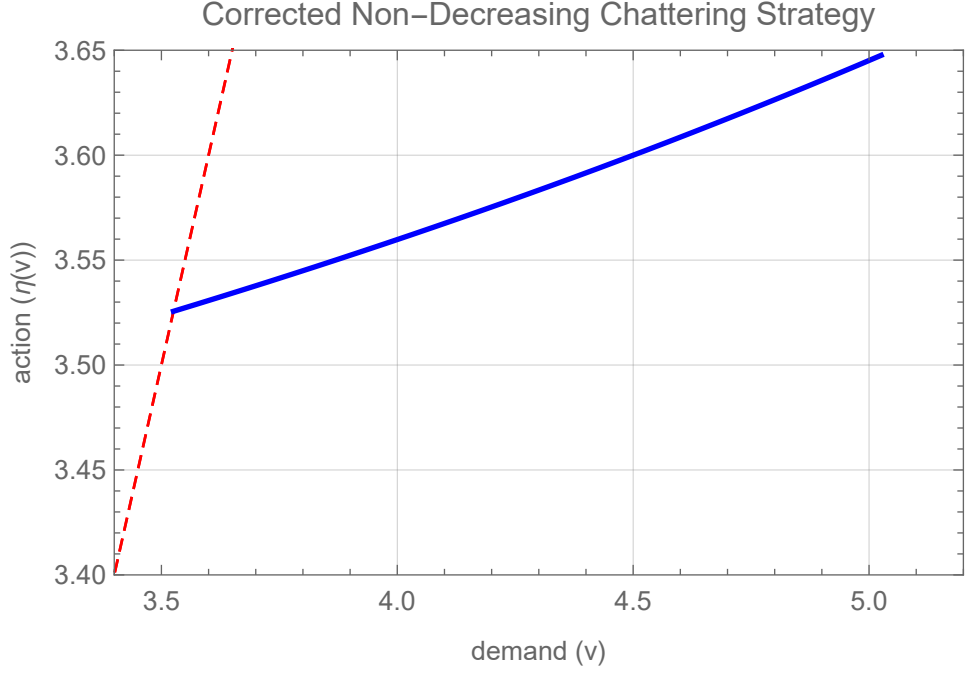
\begin{figure}[!t]
	\centering
	\begin{tikzpicture}
		\begin{axis}[
			width=0.98\columnwidth,
			height=5cm,
			axis line style={black, thin},
			tick align=inside,
			tick style={thin, black},
			grid=both,
			grid style={line width=0.4pt, draw=gray!30},
			title={Non-Decreasing Chattering Strategy},
			title style={font=\footnotesize\sffamily},
			xlabel={demand ($v$)},
			ylabel={action ($\eta(v)$)},
			label style={font=\footnotesize},
			tick label style={font=\footnotesize},
			xmin=3.4, xmax=5.2,
			ymin=3.4, ymax=3.65,
			]
			
			\addplot[
			color=red!80!black, 
			dashed, 
			line width=0.8pt,
			domain=3.4:3.65, 
			samples=2
			] {x};
			
			\addplot[
			color=blue!80!black, 
			line width=1.2pt
			] coordinates {
				(3.52562, 3.52562)
				(3.57562, 3.52905)
				(3.62562, 3.53253)
				(3.67562, 3.53604)
				(3.72562, 3.53959)
				(3.77562, 3.54318)
				(3.82562, 3.54681)
				(3.87562, 3.55048)
				(3.92562, 3.55419)
				(3.97562, 3.55794)
				(4.02562, 3.56174)
				(4.07562, 3.56558)
				(4.12562, 3.56946)
				(4.17562, 3.57338)
				(4.22562, 3.57735)
				(4.27562, 3.58137)
				(4.32562, 3.58543)
				(4.37562, 3.58954)
				(4.42562, 3.59370)
				(4.47562, 3.59791)
				(4.52562, 3.60216)
				(4.57562, 3.60647)
				(4.62562, 3.61083)
				(4.67562, 3.61524)
				(4.72562, 3.61970)
				(4.77562, 3.62421)
				(4.82562, 3.62877)
				(4.87562, 3.63339)
				(4.92562, 3.63806)
				(4.97562, 3.64279)
				(5.02562, 3.64757)
			};
			
		\end{axis}
	\end{tikzpicture}
	\caption{Chattering strategy $\eta(v)$. The identity strategy function is shown as a red dashed line.}
	\label{fig:chattering}
\end{figure}

We present some numerical examples of Algorithm \ref{alg2}. We start with an example where
resources are abundant.  Assume that $n=100$, $c_n=120$, $f$ is the
exponential pdf with mean one and $\psi(x)=0$. In this case, the equilibrium strategy function
is an AIF-1 function with $\tau_1=17.2763$.  We reduce the resources and assume that $c_n=100$.
In this case, $\tau_1=3.52562$, $P_1^*=2.2842$ and $\Fperiod{p}{1}'(\tau_1+0.001)\approx 0.84285$.
This marks an entrance of a tempered CR.  This CR terminates at $5.02562$. 
The chattering strategy $\eta(v)$ is shown in Fig. \ref{fig:chattering}.


The resurgent CRs and AIF-3 equilibrium strategy functions are quite rare.  We demonstrate an
example.  Let $n=1000$, $c_n=4000$, $\psi(x)=0.001x^2$. Assume that $f$ is a Lomax pdf with
scale 5 and $\alpha=3$. We get $\tau_1=500$, $\tau_2=999.8966$. In the right neighborhood 
$\Iperiod{p}{2}'(\tau_2+0.001) \approx -1.002927$. This marks the entrance of a resurgent CR.
We present the chattering strategy $\eta$ in Table \ref{resurgent-CR}.

Now let $c_n=2500$. Keep $f$ and $\psi$ unchanged.  The equilibrium strategy function is 
an AIF-3 function with $\tau_1=35.28$, $\tau_2=110.50$ and $\Fperiod{p}{1}'(\tau_2+0.001)=
0.0421>0$. The first local maximum of $\Iperiod{p}{2}(x)$ occurs at $\tau_3=131.57$.

\begin{table}[ht]
	\centering
	\caption{Chattering strategy function $\eta(v)$ in a resurgent CR.}
	\label{resurgent-CR}
	\setlength{\tabcolsep}{4pt} 
	\begin{tabular}{|l|ccccccc|}
		\hline
		$v$ & 999.9 & 1150 & 1250 & 1500 & 1700 & 1800 & 2000 \\ \hline
		$\eta(v)$ & 999.9 & 1000.3 & 1000.5 & 1001.0 & 1001.3 & 1001.5 & 1001.7 \\ \hline
	\end{tabular}
\end{table}


\section{Conclusions}\label{s:conclusion}
In this paper we have proposed a Bayesian game to allocate resources.
We assume that resources are granted to the players
on a smallest-request-first and all-or-nothing basis. 
For this game with two players, we have analyzed the equilibrium strategy
functions mathematically within the family of AIF
functions. We show that Nash equilibrium profiles consist of two identity functions,
two AIF functions with a common switch point, or two AIF functions with one
and three switch points, respectively.
For an $n$-player game with a large number of homogeneous players and a large $c_n$, 
we have presented a second-order mean-field approximation for its
equilibrium strategy function.  For the Gaussian
approximation, we propose an iteration-based construction algorithm for the strategy functions.
We show that the construction algorithm terminates in a finite number of iterations
and always produces a Nash equilibrium as an output.  We show the Nash equilibria of 
the Gaussian games can be AIF functions with finitely many switch points.  
We show that the Gaussian approximation model may have a chattering regime, in which
players play a continuous increasing strategy function in order to sustain a high payoff.
We have presented several numerical examples of the two-player game and the Gaussian approximation
model.

\textcolor{black}{Finally, an interesting piece of future work is to investigate
	whether the family of AIF functions is closed under best response operations of expected
	payoffs in \req{cp} or \req{constrained-Gaussian-payoff}. Another promising extension of this work is to formulate the resource allocation as a Stackelberg game, where the
	provider acts as a strategic leader who optimizes and commits to an allocation
	policy, and the users act as followers.
	Joint optimization of resource provider and user policies may lead
	to stronger results.}

\appendices
\section{Proof of \rprop{decp}, \rlem{lem1} and \rthe{main}}	\label{s:appA}
In this appendix we prove \rprop{decp}, \rlem{lem1} and \rthe{main}.
In this appendix we present detailed proofs for \rthe{main} and \rthe{global-opt}.

\begin{proof}[Proof of \rprop{decp}] 
Since we assume that the strategy function $s_{-i}$ is non-decreasing, its right-continuous lower 
inverse function $s_{-i}^{\leftarrow}(x)$ and right-continuous upper inverse function 
$s_{-i}^{\rightarrow}(x)$ are both non-decreasing. Cumulative distribution functions are also 
non-decreasing. Consequently, the composite functions $F_{-i}(s_{-i}^{\rightarrow}(x))$ and 
$F_{-i}(s_{-i}^{\leftarrow}(x)^{-})$ are non-decreasing functions of $x$. Because these two terms 
are multiplied by a negative constant ($-\frac{1}{2}$) in (6), they are non-increasing with respect 
to $x_i$. Furthermore, since $c-x$ is a strictly decreasing function of $x$, the composite 
function $F_{-i}(s_{-i}^{\rightarrow}(c-x))$ is non-increasing. For $x_i \in (c/2, c]$, $r_i(x_i)$ 
is the sum of a constant and three non-increasing terms, and therefore $r_i(x_i; s_{-i})$ is non-increasing. 
The result follows. 
\end{proof}	
	
\begin{proof}[Proof of \rlem{lem1}] To prove Statement 1, note that
\begin{align}
	\pI{i}'(v) &= 1+e^{-\lambda_{-i} v}-e^{-\lambda_{-i}(c-v)}
	-\lambda_{-i} v\left(e^{-\lambda_{-i} v}+
	e^{-\lambda_{-i}(c-v)}\right)\nonumber\\
	&\quad -\psi'(v),\label{pL'}\\
	\pI{i}''(v) &= \lambda_{-i}(\lambda_{-i} v -2)e^{-\lambda_{-i} v}-\lambda_{-i}
	(\lambda_{-i} v+2)e^{-\lambda_{-i}(c-v)}\nonumber\\
	&\quad-\psi''(v).\label{pL''}
\end{align}
To show that $\pI{i}''<0$ on $[c/2, c]$, define
\[
g(v)=v\left(1+e^{-\lambda_{-i} v}-e^{-\lambda_{-i}(c-v)}\right).
\]
Then,
\[
g''(v)=\lambda_{-i}\left((\lambda_{-i} v-2)e^{-\lambda_{-i} v}-(\lambda_{-i}
 v+2)e^{-\lambda_{-i}(c-v)} \right).
\]
On $[c/2, c]$, $v\ge c-v$ and hence, $e^{-\lambda_{-i}(c-v)}\ge e^{-\lambda_{-i} v}$. 
It follows that
\begin{align*}
g''(v)&\le \lambda_{-i}\left((\lambda_{-i} v-2)e^{-\lambda_{-i} v}-
(\lambda_{-i} v+2)e^{-\lambda_{-i} v} \right)\\
&\le -4\lambda_{-i} e^{-\lambda_{-i} v}<0.
\end{align*}
Thus, $g$ is strictly concave on $[c/2, c]$. Since $\psi(v)$ is convex, the claim follows.

Now consider Statement 2. We shall determine the sign of $\pI{i}'(c)$
in \eqref{pL'}, \ie
\[
\pI{i}'(c)=e^{-\lambda_{-i} c}(1-\lambda_{-i} c)-\lambda_{-i} c-\psi'(c).
\]
Replacing $\lambda_{-i} c$ by $y$ in the preceding equation, we define
\beq{defh}
h(y)=e^{-y}(1-y)-y-\psi'(y/\lambda_{-i}).
\eeq
Since $h(0)=1-\psi'(0)>0$ and $h(y)\to -\infty$ as $y\to\infty$, by the intermediate
value theorem \cite[Theorem 4.23, p. 93]{Rudin1976}, \req{aux} 
has at least one root on $[0, \infty)$.  Since
\[
h'(y)=(y-2)e^{-y}-1-\psi''(y/\lambda_{-i})/\lambda_{-i},
\]
it is clear that $h'(y)<0$ for all $y$ on the positive real line. Thus,
$h(y)$ is strictly decreasing.  This implies that the root of \req{aux} is unique. 
This implies that $\pI{i}'(c)>0$ if $\lambda_{-i} c<\theta$, and
$\pI{i}(c)\le 0$ if $\lambda_{-i} c\ge \theta$.  At $v=0$,
$\pI{i}(0)=2-e^{-\lambda_{-i} c}-\psi'(0)>0$, since $\psi'(0)<1$.
The unimodal property of $\pI{i}(v)$ implies that $\pI{i}'(v)=0$
has exactly one root in $[c/2, c]$ if $\lambda_{-i} c\ge \theta$, and has
no root if $\lambda_{-i} c< \theta$.

We now prove Statement 3. We differentiate \eqref{pL} twice and get
\[
\pL{i}''(x, c)=-\lambda_{-i} e^{-\lambda_{-i} (c-x)}(2+\lambda_{-i}x)-\psi''(v).
\]
Because $\lambda_{-i}>0$, $x>0$, and $\psi(x)$ is convex, the expression above
is negative.  Thus, we complete the proof of Statement 3.

For Statement 4, by the intermediate value theorem, \req{aux2} has at least one root in
$[0, \infty)$. Since the left side of \req{aux2} is decreasing, the root is unique.
Differentiating \eqref{pL} and evaluating at $c$, we obtain
\[
\pL{i}'(c,c)=e^{-\lambda_{-i}c}-\lambda_{-i}c-\psi'(c),
\]
which is negative if $\lambda_{-i}c >\phi$.

For Statement 5, from \eqref{pI}, \eqref{pT} and \eqref{pL} we have
\begin{align*}
\pI{-i}(x)-\pT{i}(x,\tau_{j+1}) &=\frac{x}{2}(F_{-i}(\tau_{j+1})-F_{-i}(x))> 0,\\
\pT{i}(x,\tau_{j+1})-\pL{i}(x,\tau_{j+1})&=\frac{x}{2}(F_{-i}(\tau_{j+1})-F_{-i}(x))>0,
\end{align*}
since $x<\tau_{j+1}$.  We have completed the proof.  
\end{proof}

\begin{proof}[\textcolor{black}{Proof of \rthe{main}}] 
To prove that $s_i$ and $s_{-i}$ obtained from Algorithm \ref{alg1} form a Nash
equilibrium, one needs to show $s_i$ and $s_{-i}$ are best responses to each other.
The analysis between \rlem{lem1} and \rthe{main} already establish that $s_{-i}$
in \req{s-i-1}, \req{s-i-2} and \req{s-i-3} are best responses to $s_i$ in \req{si}.
Here we show that $s_i$ in \req{si} is a best response to $s_{-i}$ in 
\req{s-i-1}, \req{s-i-2} and \req{s-i-3}.

First we consider the case where $\lambda_1 c\le \theta_1$ and $\lambda_2 c\le \theta_2$	
From Statements 1 and 2 of \rlem{lem1}, $\pI{i}(x)$ is increasing
on $[c/2, c]$ and $c$ is a local maximum point for $i=1, 2$.  Clearly, the two identity
functions on $[0, c]$ form a Nash equilibrium.

Now assume that $\lambda_1 c>\theta_1$ or $\lambda_2 c>\theta_2$.  In this case, $t_{i,1} \in
(c/2, c]$ is well defined. If $s_{-i}$ is of the form in \req{s-i-1}, 
for any $v < t_{i,1}$, Player $-i$ plays the identity strategy. Because $\pI{i}(x)$ 
is strictly increasing on $[0, v)$ for $v<t_{i,1}$, Player $i$'s best response is $b_i(v; s_{-i}) = v$. 
Now consider $v \ge t_{i,1}$ and action $x\in [0, v]$. \textcolor{black}{
Because the constructed $s_{-i}$ preserves the pure identity function on the interval 
$[0, t_{i,1}]$, player $i$'s payoff for any action $x \le t_{i,1}$ remains identical to 
playing against a pure identity opponent.} Thus, for $x<t_{i,1}$, player $i$'s 
payoff is increasing and has a maximum at $\pI{i}(t_{i,1})$. 
\textcolor{black}{For actions larger than $t_{i,1}$,} from Statement 5 of \rlem{lem1}
and the fact that \req{2ndsw} has no root in $(t_{i,1}, c]$, it follows that\textcolor{black}{
\beq{flatcond}
\max\Bigl\{\pT{i}(t_{i,1}, c), \sup_{x\in [t_{i,1}, c]} \pL{i}(x,c)\Bigr\} <\pI{i}(t_{i,1}). 
\eeq}
Thus, $b_i(v; s_{-i})=t_{i,1}$. We have established that $s_i$ is of the form in \req{si}.

Now consider $s_{-i}$ in the form in \req{s-i-2}. In this case, player $-i$ 
takes a flat strategy in interval $[t_{i,1}, t_{-i,2}]$ and then resumes to the
identity strategy for $t_{-i,2}< v\le c$. For $v<t_{i,1}$, the proof is identical to
that of \req{s-i-1}. We omit the replication. Consider $x\le v\in [t_{i,1}, t_{-i,2}]$.
The inequality in \req{flatcond} still holds, except that at $t_{-i,2}$ the inequality
becomes an equality. Since $p_{i}(x)\le \pI{i}(t_{i,1})$ for $x\in (t_{i,1}, t_{-i,2}]$,
it follows that $b_{i}(v; s_{-i})=t_{i,1}$ for $v\in [t_{i,1}, t_{-i,2}]$.  Finally, consider
$v\in (t_{-i,2}, c]$. From the analysis above, we have
\[
\sup_{x\in [0, t_{-i,2}]}p_i(x; s_{-i})=\pI{i}(t_{i,1}).
\]
For $v$ in $(t_{-i,2},c]$, player $-i$ adopts the identity strategy. Thus,
$p_i(x; s_{-i})$ is of the form $\pI{i}(x)$. From Statement 1 of \rlem{lem1},
$\pI{i}(x)$ is strictly concave on $[c/2, c]$.  For $x> t_{i,2}$,
$\pI{i}(x)$ is strictly decreasing, because $x$ has passed $v_i^*$.
Thus, $\pI{i}(x)<\pI{i}(t_{i,1})$
for $x> t_{i,1}$.  It follows that $b_i(v; s_{-i})=t_{i,1}$. 
The proof of $s_{-i}$ is of the form \req{s-i-3} is very similar.  We omit the details.
\end{proof}	

\section{Proof of \rlem{z-score}, \rprop{decp2}, \rprop{negdriftatc}, 
	\rprop{eta-exist-incr}, \rprop{etam-converge},
\rthe{finite}, and \rthe{global-opt}}\label{s:appB}

We first prove \rlem{z-score}.
\begin{proof}[Proof of \rlem{z-score}]
Let $N(v)$ and $D(v)$ be the numerator and the denominator of $w(v)$ defined in \req{wdef}.
Since $N(0)=c_n>0$, $N(v)\to -\infty$ as $v\to c_n$,
by the intermediate value theorem, $N(v)$ has a unique zero. Denote this zero by
$\tilde v_n$.  We claim that this zero is large and is of order $O(n)$. To see this, note
that we assume $c_n\approx n\cdot c$. For large $v$, $\mu(v) \approx\mu(\infty)$
and $\sigma(v) \approx\sigma(\infty)$.
Thus, 
\beq{xnstar}
\tilde v_n\approx c_n-(n-1)\mu(\infty)\approx n[c-\mu(\infty)].
\eeq
Now we prove Statement 2. Suppose that $v$ is of order $n$ and $v> \tilde v_n$. At this $v$,
$\mu(v)\approx\mu(\infty)$, $\sigma(v)\approx\sigma(\infty)$ and $s(v)\approx O(n\cdot x_0)$.
From \req{wdef}, we have
\beq{w-large-x}
w(v)\approx \frac{nc-s(v)-(n-1)\mu(\infty)}{\sqrt{n-1}\sigma(\infty)}
\approx\sqrt{n}\frac{c-x_0-\mu(\infty)}{\sigma(\infty)}<0.
\eeq
Now consider Statement 3.  The quotient rule
implies that 
\beq{quotient}
w'(v)=Q(v)/D(v)^2,
\eeq 
where
\beq{Q}
Q(v)=N'(v)D(v)-N(v)D'(v).
\eeq  
First, $D(v)=\sqrt{n-1}\sigma(v)\approx\sqrt{n-1}\sigma(\infty)$.
After algebraic 
manipulations, we have
\begin{align}\label{Q-temp}
Q(v)&=-\frac{\sqrt{n-1}}{\sigma(v)}\Bigl[2(1+(n-1)\mu'(v))\sigma(v)^2\nonumber\\
&\qquad+N(v)\mu'(v)(s(v)-2\mu(v))\Bigr].
\end{align}
In view of \req{xnstar}, $v$ is large and $\mu(v)$ and $\sigma(v)$ are approximately
constants.  Differentiating \eqref{mudef}, we get
\[
\mu'(v)=s(v) f(v).
\]
Thus, we can rewrite \req{Q} as
\begin{align}\label{Q2}
Q(v)&=-\frac{\sqrt{n-1}}{\sigma(\infty)}\Bigl[2\sigma(\infty)^2+2(n-1)s(v) f(v)\sigma(\infty)^2+
(cn-s(v)\nonumber\\
&\quad -(n-1)\mu(\infty))s(v) f(v)(s(v)-2\mu(\infty))\Bigr].
\end{align}
There are three terms in the square brackets in the preceding equation.
The first term is constant and is positive. Since $s(v)$ is of order $n$, 
the second term in the square bracket in \eqref{Q2} is positive and is
of order $O(n^2 f(n))$. The third term is negative and its magnitude is 
of order $O(n^3 f(n))$. Both the second term and the third term approach
to zero according to \rassum{pdfshape}.  Thus, 
\beq{Q-order}
Q(v)\approx -2\sqrt{n-1}\sigma(\infty)<0.
\eeq
Substituting \req{Q-order} into \req{quotient}, we have
\[
w'(v)\approx -2n^{-1/2}/\sigma(\infty)<0.
\]
Thus, we have proven Statement 3.
\end{proof}
 
\begin{proof}[Proof of \rprop{decp2}]
We first consider $\Iperiod{w}{k}(v)$. Let $\Iperiod{N}{k}(v)$ and $\Iperiod{D}{k}(v)$
be the numerator and the denominator of $\Iperiod{w}{k}(v)$.
Note that 
\begin{align}
	\Iperiod{N}{k}'(v) &= -1-(n-1)\Iperiod{\mu}{k}'(v) \nonumber\\
	&=-1 -(n-1)v f(v).\label{negdrift}
\end{align}	
Clearly $\Iperiod{N}{k}'(v)<0$, and thus $\Iperiod{N}{k}(v)$ is decreasing.
Suppose that $\Iperiod{N}{k}(v)$ is non-negative. It would be sufficient to guarantee that
$\Iperiod{w}{k}(v)$ is decreasing, if one can show that
$\Iperiod{D}{k}(v)$ is increasing.  On the other hand, if $\Iperiod{N}{k}(v)$ is negative,
one needs to examine the derivative of $\Iperiod{w}{k}(v)$ in order to guarantee that it
is decreasing.  Thus, we consider two cases.  In the first case, we consider $v<\tilde v_n$.
In this case, $\Iperiod{N}{k}(v)\ge 0$. 
Differentiate $\Iperiod{D}{k}(v)$, we obtain
\beq{D'}
\Iperiod{D}{k}'(v)=\sqrt{n-1}\Iperiod{\sigma}{k}'(v)=\sqrt{n-1}
\left(\frac{s(v)-2\Iperiod{\mu}{k}(v)}{2\Iperiod{\sigma}{k}(v)}\right)\Iperiod{\mu}{k}'(v).
\eeq
Since $\Iperiod{\mu}{k}'(v)>0$, $\Iperiod{D}{k}'(v)>0$ if we can show that
\beq{x-2mu}
s(v)> 2\Iperiod{\mu}{k}(v).
\eeq 
To prove \req{x-2mu}, since $v$ is in an identity interval and $s(v)=v$,
it suffices to show
\beq{goal}
\frac{v}{2}-\int_0^v t f(t)\,dt \ge 0.
\eeq
Let 
\[
H_1(v)=\frac{v F(v)}{2}-\int_0^v t f(t)\,dt.
\]
Note that $v/2 \ge v F(v)/2$.  Thus, $H_1(v)>0$ implies that \req{goal} holds.
Now $H_1(0)=0$.  Our goal is to show that $H_1(v)$ is increasing in $v$.
Differentiate and get
\[
H_1'(v)=\frac{1}{2}(F(v)-v f(v)).
\]
Now 
\beq{xfx}
F(v)-v f(v)=\int_0^v f(t)\,dt-\int_0^v f(t)\,dt=\int_0^v[f(t)-f(v)]\,dt
\eeq
The last quantity in \req{xfx} is non-negative since we assume that $f$ is non-increasing
in \rassum{pdfshape}. Thus, $H_1(v)$ is increasing. 

Now we consider the second case where $v\ge \tilde v_n$. In this case, $N(v)<0$ and we
examine the derivative of $\Iperiod{w}{k}'(v)$, which has the same sign as
\[
\Iperiod{Q}{k}(v)=\Iperiod{N}{k}'(v)\Iperiod{D}{k}'(v)-\Iperiod{N}{k}(v)\Iperiod{D}{k}'(v).
\]
After algebraic manipulations, we have
\begin{align}
\label{Q1}
\Iperiod{Q}{k}(v) &=-\frac{\sqrt{n-1}}{\Iperiod{\sigma}{k}(v)}
\Bigl[2(1+(n-1)\Iperiod{\mu}{k}'(v))\sigma(v)^2\nonumber\\
&\quad +\Iperiod{N}{k}(v)\Iperiod{\mu}{k}'(v)
(v-2\Iperiod{\mu}{k}(v))\Bigr].
\end{align}
In view of \req{xnstar}, $v$ is large and $\Iperiod{\mu}{k}(v)$ and 
$\Iperiod{\sigma}{k}(v)$ are approximately
constants.  Substituting $\Iperiod{\mu}{k}'(v)=v f(v)$ into \eqref{Q1}, we rewrite \eqref{Q1} as
\begin{align*}
\Iperiod{Q}{k}(v)&=-\frac{\sqrt{n-1}}{\Iperiod{\sigma}{k}(\infty)}\Bigl[
2\Iperiod{\sigma}{k}(\infty)^2+2(n-1)v f(v)\Iperiod{\sigma}{k}(\infty)^2+\nonumber\\
&\quad(cn-v-(n-1)\Iperiod{\mu}{k}(\infty))v f(v)(v-2\Iperiod{\mu}{k}(\infty))\Bigr].
\end{align*}
There are three terms in the square brackets in the preceding equation.
The first term is constant and is positive. The second term is positive and is
of order $O(n^2 f(n))$. The third term is of order $O(n^3 f(n))$. Both terms approach
to zero according to \rassum{pdfshape}.  Thus, 
\[
Q(v)\approx -2\sqrt{n-1}\sigma(\infty)<0.
\]
We have completed the proof of Statement 1 for the case of $\Iperiod{w}{k}(v)$.
The proof that $\Fperiod{w}{k}'(v)< 0$ is very similar.  We omit the details. 

Since $\Phi$ is increasing, Statement 2 follows from Statement 1.
\end{proof}
	
\begin{proof}[Proof of \rprop{negdriftatc}]
Assume, for the sake of contradiction, that the supremum of the equilibrium 
action space reaches the capacity limit, such that $\sup_{v \in [0, \bar{v}]} s(v) = c_n$
for some $\bar v\in [0, c_n]$. 
Since the strategy function $s(v)$ is monotonically non-decreasing, it follows that 
as $v \to \bar{v}$, the action $s(v) \to c_n$. Recall that the 
standard $z$-score governing the normal approximation of the success 
probability is given in \req{wdef}. From Statement 2 of \rlem{z-score},
$w(v)$ is negative when $v$ is large.  Particularly, for $s(v)$ in the left neighborhood
of $c_n$, we have
\[
\lim_{s \to c_n^-} w(v) = \frac{c_n - c_n - (n-1)\mu(\bar v)}{\sqrt{n-1}\sigma(\bar v)} 
= -\sqrt{n-1} \frac{\mu(\bar v)}{\sigma(\bar v)}.
\]
Clearly $w(\var v)$ is strictly negative and is of order $O(n)$. It follows that
$\Phi(w(\bar v))\approx 0$.  Thus, $p(v)$ defined in \req{pdef} at $v=\bar v$ is
\[
p(\bar v)\approx -\psi(c_n)<0.
\]
This strict negativity indicates that any player with a value $v$ close to $\bar{v}$ 
would strictly prefer to deviate by reducing their action $s$, contradicting the 
assumption that $s(v) \to c_n$ is an equilibrium best response.
\end{proof}	

\begin{proof}[Proof of \rprop{eta-exist-incr}] 
We differentiate \req{sustain-rec-high} to obtain
\beq{etap}
\eta'(v)=-\frac{\eta(v)\phi(w_\eta(v))w'(v)}{\Phi(w_\eta(v))-\psi'(\eta(v))},
\eeq
where $\phi$ is the pdf of a standard normal random variable.
Differentiating \eqref{mueta}, we have
\beq{mup2}
\mu'(v)=\eta(v) f(v).
\eeq
Differentiating \eqref{sigmaeta} and rearranging terms, we obtain
\beq{sigmap}
\sigma'(v)=\frac{v-2\mu(v)}{2\sigma(v)}\mu'(v).
\eeq
Eqs. \req{etap}, \req{mup2} and \req{sigmap} form a system of ordinary differential equations with
initial condition 
\begin{align*}
&(\eta(\tau),\mu(\tau),\sigma(\tau))=\\
&\left(\tau, \int_0^\tau s(t)f(t)\,dt,
\sqrt{\int_0^\tau s(t)^2 f(t)\,dt-\mu(\tau)^2}\right).
\end{align*}
A CR can be entered at the end of an identity interval, or at the end of a flat interval.
In this proof we assume that a CR is entered at the end of the $k$-th identity interval, \ie
$\tau=\tau_{2k-1}$. The proof for the other case is similar.
At the entry point $\tau$, we must verify that the solution $\eta(\tau)$ is valid, \ie, 
that the slope $\eta'(\tau)$ represents a feasible mixture of identity and flat behaviors. 
Let $\Iperiod{p}{k}(v)$ (defined in \req{defpk}) and $\Fperiod{p}{k}(v)$ denote the potential 
unconstrained payoffs if the player were to continue with a pure identity or pure flat 
strategy, respectively. The left hand side of \req{sustain-rec-high} is the unconstrained
payoff if strategy $\eta(v)$ is used for $v\ge \tau$. We relate the derivatives
of the unconstrained payoff associated with $\eta(v)$ with $\Iperiod{p}{k}'(\tau)$ and
$\Fperiod{p}{k}'(\tau)$. From \eqref{defpkF} and \eqref{defwkF} we have
\begin{align*}
	\Fperiod{p}{k}(v) &= \tau \Phi(\Fperiod{w}{k}(v))-\psi(\tau) \\
	\Fperiod{w}{k}(v) &=\frac{c_n-\tau-(n-1)\Fperiod{\mu}{k}(v)}{\sqrt{n-1}\Fperiod{\sigma}{k}(v)}.
\end{align*}
From \eqref{muidentity}, \req{muflat} and \eqref{mueta}, $\Iperiod{\mu}{k}(\tau)=\Fperiod{\mu}{k}(\tau)
=\mu(\tau)$.  Similarly, from \req{sigmaident}, \eqref{sigmaflat} and \eqref{sigmaeta}, we have
$\Iperiod{\sigma}{k}(\tau)=\Fperiod{\sigma}{k}(\tau)=\sigma(\tau)$. These imply that
$\Iperiod{w}{k}(\tau)=\Fperiod{w}{k}(\tau)=w_\eta(\tau)$. Denote $\Iperiod{w}{k}(\tau)$ by $W$.
Denote $\Fperiod{w}{k}'(\tau^+)$ by $\Omega$ and $\sqrt{n-1}\Fperiod{\sigma}{k}(\tau)$ by $D$.
Note that $\Omega$ represents the rate of change of the the z-score of a normal distribution 
independent of the player's own marginal action change.  Differentiating, one can verify that 
\begin{align*}
	\Iperiod{w}{k}'(\tau^+) &= \Omega-\frac{1}{D} \\
	w_\eta'(\tau^+) &= \Omega-\frac{\eta'(\tau)}{D}.
\end{align*}
Substituting these into the payoffs, one obtain
\begin{align}
\Fperiod{p}{k}'(\tau^+) &= \tau \phi(W)\Omega \label{pFprime}\\
\Iperiod{p}{k}'(\tau^+) &= (\Phi(W)-\psi'(\tau))+\tau\phi(W)\left(\Omega-
\frac{1}{D}\right) \label{pIprime}\\
p_\eta'(\tau^+) &= \eta'(\tau)(\Phi(W)-\psi'(\tau))+\tau\phi(W)
\left(\Omega-\frac{\eta'(\tau)}{D}\right).	\label{petaprime}
\end{align}
From \eqref{pFprime} and \eqref{pIprime}, we establish that
\beq{der-diff}
\Iperiod{p}{k}'(\tau^+)-\Fperiod{p}{k}'(\tau^+)=\Phi(W)-\psi'(\tau)-\frac{\tau\phi(W)}{D}.
\eeq
For all $v$ including $\tau$ in a CR, $p_\eta'(v)=0$, since $p_\eta(v)$ is maintained at a constant. 
Thus, the left side of \eqref{petaprime} is equal to zero. 
Substituting \req{der-diff} and \eqref{pFprime} into the right side of
\eqref{petaprime}, we get \req{etap2}.
The gradient conflict condition in \req{grad-contra} implies that $0<\eta'(\tau)\le 1$.
This proves that the trajectory enters the Chattering Regime with a valid, strictly 
increasing slope that lies strictly between the slopes of the flat (0) and identity (1) strategies.

We now verify the conditions of the Picard–Lindel\"{o}f Theorem \cite[Thm 5.15, p. 119]{Rudin1976}.
The right hand side of the system in \req{etap}, \req{mup2} and \req{sigmap}
is a vector-valued real function. Note that $\Phi$, $\psi$, $f$ and $\psi$
are analytic (or piece-wise analytic according to \rassum{analyticity}).
Under \rassum{marginal}, the denominator of \req{etap} is greater than some $\epsilon$, where
$\epsilon>0$. Provided that the pdf $f$ is not degenerate, $\sigma(v)$ remains positive, 
preventing the denominator of \req{sigmap} from vanishing.  Thus, the
vector-valued function is continuous. The derivatives of the right-hand side with 
respect to $\eta$, $\mu$, and $\sigma$ are well-defined and bounded because the 
denominators are bounded away from zero (as established above).
The system has no critical points within the regime that would cause a singular 
Jacobian matrix, ensuring the Lipschitz constant is finite over the compact interval 
$[0, c_n]$. By the Picard-Lindel\"{o}f Theorem, the system in \req{etap}, \req{mup2} 
and \req{sigmap} has a unique solution. Since the solution to the ODE system is unique 
(as proven via Picard-Lindelöf), all convergent subsequences must converge to the same 
limit $\eta(v)$. Therefore, the entire sequence $\{\eta_m(v)\}$ converges uniformly to $\eta(v)$.

Now we prove that $\eta(v)$ is strictly increasing. Assume that $c_n\approx nc$ and $n$ is large, 
\[
w_\eta(v)\approx \sqrt{n}\frac{c-\mu(v)}{\sigma(v)}.
\]
Since $\sqrt{n}$ is large, $\Phi(w_\eta(v))$ behaves like a step function
creating three regimes for the term $\Phi(w_\eta(v))$ in \req{sustain-rec-high}.
In the first regime where $\mu(v)<c$, $\Phi(w_\eta(v))\approx 1$.
Thus,
\[
\eta(v)\cdot 1-\psi(\eta(v))\approx P^*,
\]
which implies that $\eta(v)$ is a constant in $v$ in this regime. This implies that
$\eta'(v)=0$ and $\eta$ is non-decreasing.

In the second regime where $\mu(v)>c$, $\Phi(w_\eta(v))\approx 0$. In this case, \req{sustain-rec-high}
simplifies to $-\psi(\eta(v))\approx P_k^*$, which has no solution. Thus, a CR can not occur
in this case.  Finally, consider the critical regime where $\Phi(w_\eta(v))\in (0, 1)$.
Now examine the sign of $\eta'(v)$ in \req{etap}.
Differentiating \eqref{weta} and applying large $n$ analysis, we obtain
\[
w_\eta'(v)\approx -\sqrt{n}\frac{\mu'(v)}{\sigma(v)}.
\]
Since $\mu'(v)=\eta(v)f(v)>0$, it is clear that $w_\eta'(v)$ is negative.
From \rassum{marginal}, the denominator of the right side of \req{etap} is greater than 
some $\epsilon$, where $\epsilon>0$. Thus, $\eta'(v)>0$ and $\eta$ is increasing.
\end{proof}

\begin{proof}[Proof of \rprop{etam-converge}] 
	The sequence of strategy functions $\{\eta_m(v)\}$ is constructed such that for each 
	grid point $v_j = \tau + j\delta$, the value $\eta_m(v_{j+1})$ is determined by the iteration
	$\eta_m(v_{j+1}) = \eta_m(v_j) + \eta'(v_j)\delta$. This iteration is precisely the Forward Euler 
	discretization of the system of ordinary differential equations \req{etap}--\req{sigmap}. 
	
	According to \rprop{eta-exist-incr}, the right-hand side of the ODE system, denoted as 
	$\mathcal{G}(v, \eta, \mu, \sigma)$, is Lipschitz continuous because its components 
	are piecewise analytic and the denominator $\Phi(w) - \psi'(\eta)$ is bounded away 
	from zero. Let $L_{\mathcal{G}}$ be the Lipschitz constant of $\mathcal{G}$. 
	A standard result in numerical analysis for the Forward Euler method on a compact 
	interval $[\tau, c_n]$ ensures that the global truncation error at the grid points is bounded by:
	\beq{truncation-error}
		| \eta_m(v_j) - \eta(v_j) | \le C_G \cdot \delta, \quad \forall j=0, \dots, m,
	\eeq
	where $C_G$ is a constant independent of $j$ and $v$ \cite{Ascher1998Book}.
	
	Now consider any $v \in [\tau, c_n]$. Let $v_j$ be the grid point such 
	that $v_j \le v < v_{j+1}$. By applying the triangle inequality, the error 
	at $v$ is bounded by:
	\beq{triangle-ineq}
		| \eta_m(v) - \eta(v) | \le | \eta_m(v) - \eta_m(v_j) | + | \eta_m(v_j) - \eta(v_j) | 
		+ | \eta(v_j) - \eta(v) |.
	\eeq
	We evaluate each term as follows:
	\begin{itemize}
		\item By construction in \req{etam}, $\eta_m(v)$ consists of segments with slopes 0 or 1. 
		Thus, $\eta_m(v)$ is Lipschitz continuous with constant $L=1$, 
		implying $| \eta_m(v) - \eta_m(v_j) | \le 1 \cdot |v - v_j| \le \delta$.
		\item As established in \req{truncation-error}, the grid point error 
		is $| \eta_m(v_j) - \eta(v_j) | \le C_G \delta$.
		\item From \rprop{eta-exist-incr}, $\eta(v)$ is strictly increasing with a derivative $0 < \eta'(v) < 1$. 
		Thus, $\eta(v)$ is also Lipschitz continuous with constant $L=1$, 
		implying $| \eta(v_j) - \eta(v) | \le 1 \cdot |v_j - v| \le \delta$.
	\end{itemize}
	
	Combining these results and substituting into \req{triangle-ineq}, we obtain:
	\[
		| \eta_m(v) - \eta(v) | \le \delta + C_G \delta + \delta = (2 + C_G) \delta = O(\delta).
	\]
	Since the bound $(2 + C_G) \delta$ is independent of the location of $v$, it follows that:
	\[
		\lim_{m \to \infty} \sup_{v \in [\tau, c_n]} | \eta_m(v) - \eta(v) | = 0,
	\]
	which proves uniform convergence of the sequence $\{\eta_m(v)\}$ to the chattering strategy $\eta(v)$.
\end{proof}

\begin{proof}[Proof of \rthe{finite}]
We prove this theorem by contradiction.  Assume that the algorithm generates
an infinite sequence of switch points $0=\tau_0<\tau_1<\ldots$ converging to
an accumulation point $\tau^*\le c_n$. If $\tau^*<c_n$, the strategy function
on $[\tau^*, c_n]$ may be a flat function, or a mixed strategy function $\eta(v)$.
As $k \to \infty$, the length of the intervals $\delta_k = \tau_k - \tau_{k-1}$ approaches zero.
Consider the strategy function $s(v)$ constructed by the algorithm.
For $v \in [\tau_{2k}, \tau_{2k+1})$ (identity interval), $s(v) = v$.
For $v \in [\tau_{2k-1}, \tau_{2k})$ (flat interval), $s(v) = \tau_{2k-1}$.
In the limit as $k \to \infty$, for any $v$ in the neighborhood 
of $\tau^*$, we have $|s(v) - v| \le \delta_k \to 0$.
Thus, the strategy function $s(v)$ converges uniformly to the 
identity function $s_{\infty}(v) = v$ on the domain approaching $\tau^*$ from the left.
Let $p_{\infty}(x)$ denote the hypothetical payoff function when players 2 to $n$ 
employ the pure identity strategy $s(v)=v$ in the left neighborhood of $\tau^*$.
Since the payoff functions $p_{k,I}(x)$ and $\Fperiod{p}{k}(x)$ are integral functionals 
of the strategy $s(v)$ (via the moments $\Fperiod{\mu}{k}(x)$ and 
$\Fperiod{\sigma}{k}(x)$), the uniform 
convergence of $s(v) \to v$ implies the uniform convergence of the payoff 
functions and their derivatives to the limit function:
\begin{align*}
	&\lim_{k \to \infty} \Iperiod{p}{k}(x) = \lim_{k \to \infty} \Fperiod{p}{k}(x) 
	=p_{\infty}(x),\\
	&\lim_{k \to \infty} \Iperiod{p}{k}'(x) = \lim_{k \to \infty} \Fperiod{p}{k}'(x)
	=p'_{\infty}(x)
\end{align*}
for $x$ in the left neighborhood of $\tau^*$.	
	
The switching conditions of Algorithm 2 impose constraints on the 
derivatives of the payoff functions at the switch points.
\begin{itemize}
	\item \textbf{End of identity ($\tau_{2k-1}$):} The identity interval ends 
	because the payoff ceases to increase. Thus, $\Iperiod{p}{k}'(\tau_{2k-1}) \le 0$ 
	(or crosses zero from positive to negative).
	\item \textbf{End of flat ($\tau_{2k}$):} The flat interval ends because the 
	payoff recovers to the previous peak $P_k^*$. For the function to cross $P_k^*$ 
	from below, we must have $\Fperiod{p}{k}'(\tau_{2k}) > 0$.
\end{itemize}
An infinite sequence of switches implies that in any $\epsilon$-left-neighborhood of $\tau^*$, 
the derivative of the payoff function flips sign infinitely many times 
(from below zero to above zero and back).

However, the limit function $p_{\infty}(x)$ is analytic (under \rassum{analyticity}). 
An analytic function on a compact interval cannot have a derivative that 
oscillates infinitely many times around zero (unless it is constant, which 
is ruled out by the problem parameters) \cite[Theorem 8.5, p. 177]{Rudin1976}.
There exists a neighborhood $(\tau^* - \delta, \tau^*)$ wherein $p'_{\infty}(x)$ 
has a constant sign (or is isolated zero).
\begin{itemize}
	\item If $p'_{\infty}(x) > 0$, then for sufficiently large $k$, $\Iperiod{p}{k}'(x) > 0$, 
	preventing the termination of identity intervals.
	\item If $p'_{\infty}(x) < 0$, then for sufficiently large $k$, $\Fperiod{p}{k}'(x) < 0$, 
	preventing the recovery of flat intervals.
\end{itemize}
This contradiction proves that an infinite sequence of switch points is 
impossible. The algorithm must terminate in a finite number of iterations.
\end{proof}

\begin{proof}[Proof of \rthe{global-opt}]
We shall prove the following three claims.  Clearly, the following three claims ensure
that the output of Algorithm 2 is globally optimal to player 1, if all other players play it.
\begin{description}
\item[Claim 1]{\bf [Optimality of AIF strategies]} Any AIF strategy function constructed
by Algorithm \ref{alg2} in AIF regime is a best response if all other players adopt it.
\item[Claim 2]{\bf [Optimality of CR]} Strategy function $\eta$ constructed in a CR
is a best response if all other players adopt it.
\item[Claim 3]{\bf [Global consistency]} No player with demand values in a CR has an 
incentive to deviate back to an AIF strategy function obtained in the AIF regime.
\end{description}

First, we prove claim 1.
Suppose that Algorithm \ref{alg2} produces an AIF function $s$ in the AIF regime as its output.
Suppose that players 2 to $n$ adopt function $s$. We prove that $s$ is player 1's best response.
Suppose that the AIF regime ends at point $\tau$ and that there are $k$ identity intervals and 
$k$ or $k-1$ flat intervals.
For demand value $v$, player 1's best response is
\begin{align}
\textcolor{black}{b(v; s)} &=\min_{0\le x\le c_n}\left\{\min\{x, v\} \Phi(w(x))-\psi(x)\right\} \nonumber \\
&= \min_{0\le x\le v}\left\{x \Phi(w(x))-\psi(x)\right\},\label{br}
\end{align}
since $w(x)$ is decreasing in games with large $n$. Suppose $v\in [\tau_{2j-2}, \tau_{2j-1}]$ for 
any $1\le j\le k$. Players 2 to
$n$ take an identity strategy.  In this case, player 1's expected payoff is $\Iperiod{p}{j}(x)$
for action $x$. Since $\Iperiod{p}{j}(x)$ is increasing on $[\tau_{2j-2},\tau_{2j-1}]$, it is clear
from \eqref{br} that $\textcolor{black}{b(v; s)}=v$.  Now suppose that $v\in (\tau_{2j-1}, \tau_{2j}]$.
Players 2 to $n$ take a capping strategy.  Thus, player 1's payoff function is $\Fperiod{p}{j}(x)$.
According to Algorithm \ref{alg2}, $\Fperiod{p}{j}(x)\le \Iperiod{p}{j}(\tau_{2j-1})$. This implies
that $\textcolor{black}{b(v; s)}=\tau_{2j-1}$.  It follows that 
\[
\textcolor{black}{b(v; s)}=\left\{\begin{array}{ll}
	v, & v\in [\tau_{2j-2}, \tau_{2j-1}], \\
	\tau_{2j-1}, & v \in [\tau_{2j-1}, \tau_{2j}).\end{array}\right.
	\]
Clearly in the $j$-th pair of identity/flat intervals, $\textcolor{black}{b(v; s)=s(v)}$.  We also note
that the construction in Algorithm \ref{alg2} ensures that $\{P_j^*, 1\le j\le k\}$ is
a strictly increasing sequence.  Thus, for any $v\in [\tau_{2k-1}, c_n]$
\beq{AIF-max}
\max_{0\le x\le \tau_{2k-1}} p(x)=P_k^*.
\eeq

Now we prove claim 2. Suppose that player 1's demand $v$ is in a CR.
From Eqs. \req{sustain-rec-high} to \eqref{sigmaeta}, we see that player 1's 
payoff is equal to $P_k^*$, if all players play strategy $\eta$. From \req{AIF-max}
this payoff is strictly better than those correspond to any action in $[0, \tau_{2k-1})$.
Thus, player 1 has no incentives to deviate from $\eta(v)$.

Finally, the argument for claim 3 is similar to that of claim 2.  We omit the redundancy.
\end{proof}

\bibliography{bibdatabase}
\bibliographystyle{IEEEtran}

\end{document}